\documentclass[12pt]{article}

\usepackage{fullpage}
\usepackage{amsmath}
\usepackage{amssymb}
\usepackage{amsthm}
\usepackage{hyperref}

\newtheorem{definition}{Definition}
\newtheorem{theorem}{Theorem}
\newtheorem{lemma}{Lemma}
\newcommand\doi[1]{\url{https://doi.org/#1}}

\usepackage{xcolor}
\definecolor{Carmine}{HTML}{9D000C}
\definecolor{Sapphire Blue}{HTML}{006BA6}

\usepackage{booktabs}

\usepackage{enumitem}
\setlist{noitemsep}

\usepackage{stmaryrd}

\usepackage{mathtools}

\SetSymbolFont{stmry}{bold}{U}{stmry}{m}{n}

\usepackage{tikz}

\usepackage{algorithm}
\usepackage[noend]{algpseudocode}

\newcommand\return[1]{\State\Return{#1}}
\newcommand\assign[2]{\State{{#1} $\gets$ {#2}}}
\renewcommand\Call[2]{\textproc{#1}%
(#2)}
\newcommand\funname[1]{\normalfont\textsc{#1}}

\usepackage{wrapfig}
\usepackage{subcaption}

\newcommand\sfact[1]{\ensuremath{\text{\normalshape¡}{#1}!}}
\newcommand\littleO[1]{o\!\left(#1\right)}
\newcommand\bigO[1]{O\!\left(#1\right)}
\newcommand\indicator[1]{\boldsymbol{1}_{\{#1\}}}
\DeclareMathOperator\VM{\mathrm{VM}}
\DeclareMathOperator\decomp{\mathrm{decomp}}
\newcommand\PP{\mathbb{P}}
\newcommand\ZZ{\mathbb{Z}}

\title{Asymptotic analysis and efficient random sampling of directed ordered
acyclic graphs}

\author{
  Martin P\'epin%
  \footnote{Université Caen Normandie, ENSICAEN, CNRS, Normandie Univ, GREYC
  UMR 6072, F-14000 Caen, France. (\texttt{martin.pepin@unicaen.fr}).}%
  \and%
  Alfredo Viola%
  \footnote{Casa de Investigadores Científicos La Comarca. Montevideo, Uruguay
  (\texttt{alfredo.viola@gmail.com}).}%
}

\begin{document}
  \maketitle

  \begin{abstract}
    Directed acyclic graphs (DAGs) are directed graphs in which there is no path
    from a vertex to itself.
    DAGs are an omnipresent data structure in computer science and the problem
    of counting the DAGs of given number of vertices and to sample them
    uniformly at random has been solved respectively in the 70's and the 00's.
    In this paper, we propose to explore a new variation of this model where
    DAGs are endowed with an independent ordering of the out-edges of each
    vertex, thus allowing to model a wide range of existing data structures.

    We provide efficient algorithms for sampling objects of this new class, both
    with or without control on the number of edges, and obtain an asymptotic
    equivalent of their number.
    We also show the applicability of our method by providing an effective
    algorithm for the random generation of classical labelled DAGs with a
    prescribed number of vertices and edges, based on a similar approach.
    This is the first known algorithm for sampling labelled DAGs with full
    control on the number of edges, and it meets a need in terms of
    applications, that had already been acknowledged in the literature.
  \end{abstract}

\section{Introduction}\label{sec:intro}

Directed Acyclic Graphs (DAGs for short) are directed graphs in which there is
no directed path (sequence of incident edges) from a vertex to itself.
They are an omnipresent data structure in various areas of computer science and
mathematics.
In concurrency theory for instance, scheduling problems usually define a partial
order on a number of tasks, which is naturally encoded as DAG via its Hasse
diagram~\cite{CMPTVW2010,CEH2019}: each task corresponds to a vertex in the
graph and task dependencies are materialised by directed edges.
Scheduling then corresponds to finding a good topological order on this graph.
Natural question such as counting or sampling the schedulings of a program are
studied in this context for the purpose of random testing~\cite{GS2005}.
DAGs also appear as the result of the compression of some tree-like structures
such as XML documents~\cite{BLMN2015}.
In functional programming in particular, this happens at the memory layout level
of persistent tree-like values, where the term ``hash-consing'' has been coined
to refer to this compression~\cite{Goto1974}.
Computer algebra systems also make use of this idea to store their symbolic
expression~\cite{Ershov1958}.
These tree use cases leverage the fact that actual memory gains can be obtained
by compacting trees, which has been quantified in~\cite{FSS1990} and has
motivated the study of compacted trees in the recent years~\cite{EFW2021,
GW2024}.
Finally, complex histories, such as those used in version control systems (see
Git for instance~\cite[p.~17]{git2018}) or genealogy ``trees'' are DAGs as well.

Most of the applications presented here actually require to add some more
structure on the space of DAGs in order to faithfully model the objects at play,
which is the main motivation of the present article.
We first give some background on the combinatorics of DAGs and then expand on
our contributions.

\subsection{Background on DAGs}

Two different models of DAGs have received a particular interest:
\emph{labelled} DAGs and \emph{unlabelled} DAGs.
The most obvious one is the labelled model, in which one has a set~$V$ of
vertices (often~$\llbracket 1; n \rrbracket$) connected by a set of edges~$E
\subseteq V \times V$.
The term \emph{labelled} is used because the vertices can be distinguished
here, they can be assigned labels.
On the other hand, unlabelled DAGs are the quotient set obtained by considering
labelled DAGs up to relabelling, that is to say up to a permutation of their
vertices (which is reflected on the edges).
These two types of objects serve a different purpose, the former represents
relations over a given set whereas the latter represents purely structural
objects.
From a combinatorial point of view, a crucial difference between the two models
is that one has to deal with symmetries when enumerating unlabelled DAGs which
makes the counting and sampling problem significantly more involved.

\paragraph{Counting}

The problem of counting DAGs has been solved in early 70's by Robinson and
Stanley using different approaches.
In~\cite{Robinson1970}, Robinson exhibits a recursive decompositions of labelled
DAGs leading to a recurrence satisfied by the numbers~$A_{n,k}$ of DAGs with~$n$
vertices including~$k$ sources (vertices without any incoming edge).
He later reformulates those results in terms of a new kind of generating
functions, now called \emph{graphical generating functions}
in~\cite{Robinson1973}, and also obtains the asymptotic number of size~$n$ DAGs.
Around the same time, Stanley also used a generating function approach
in~\cite{Stanley1973} obtained the same results by deriving identities of the
chromatic polynomial.
Robinson also solves the unlabelled case starting from the same ideas but using
Burnside's lemma and cycle index sums to account for the symmetries of these
objects.
He provides a first solution in~\cite{Robinson1970} and makes it more
computationally tractable in~\cite{Robinson1977}.
In the 90's, Gessel generalised those results, also using the graphical
generating function framework in~\cite{Gessel1995,Gessel1996} to take into
account more parameters and count DAGs by vertices and edges, but also sinks and
sources.

\paragraph{Random sampling}

From the point of view of uniform random generation, the recursive decomposition
exhibited by Robinson in~\cite{Robinson1973} is interesting as it is amenable to
\emph{the recursive method} pioneered by Nijenhuis and Wilf in~\cite{NW1978}.
This yields a polynomial time algorithm for sampling uniform DAGs with~$n$
vertices.
The analysis of this algorithm has been done in~\cite{KM2015} but it had been
acknowledged earlier in~\cite{MDB2001} although the article proposes an
alternative solution.
Both~\cite{KM2015} and~\cite{MDB2001} also offer a Markov chain approach to the
random sampling problem and an interesting discussion on the pros and cons of
both approaches is given in~\cite{KM2015}.
Remote from the field of combinatorics, the random generation of DAGs is also an
active topic in the area of applied statistics and Bayesian inference.
In this context, DAGs encode a relevant structure in a collection of random
variables and the problem of interest is to sample DAGs from a particular
distribution related to those random variables.
To this end, authors resort both to Monte Carlo Markov Chains
approaches~\cite{KSM2022, KM2017} and methods similar to what is referred to as
the \emph{recursive method} in combinatorics~\cite{TVK2020}.
An important point in~\cite{KM2017} is better performance can be achieved by
using a combination of both approaches, in particular by exploiting the
combinatorial properties of DAGs.
Notable is that sampling from the uniform distribution is tackled as a
particular case in~\cite{TVK2020} and solved with the asymptotically
optimal~$O(n^2)$ complexity at the expense of a~$O(n^3)$ pre-processing step.

Unfortunately, to our knowledge, no efficient uniform random generator of
unlabelled has been found yet.
Moreover, unlike in the labelled case, the method derived by Robinson to exhibit
the number of unlabelled DAGs cannot be easily leveraged into a random sampler
as they make extensive use of Burnside's lemma.

Another interesting question is that of controlling the number of edges in
those random samplers.
Indeed, sampling a uniform DAG with a prescribed number of vertices and edges
cannot be achieved using the Markov chain approach as it constrains the chain
too much, and the formulas of Gessel are not amenable to this either.
In~\cite[\textsection~7]{KM2015}, the authors provide an interesting discussion
on which kind of restrictions can be made on DAGs with the Markov chain
approach.
They address in particular the case of bounding the number of edges and
highlight that the Monte Carlo Markov Chain approach fails when the desired
number of edges is too low, thus advocating for having precise combinatorial
enumerations.

\subsection{Contributions}

In the present paper, we propose to study an alternative model of DAGs, which we
call Directed Ordered Acyclic Graphs (DOAGs), and which are enriched with
additional structure on the edges.
More precisely, a DOAG in an unlabelled DAG where (1) set of outgoing edges of
each vertex is totally ordered and (2) the sources are totally ordered as well.
This \emph{local} ordering of the outgoing edges allows to capture more
precisely the structure of existing mathematical objects.
For instance, the compressed formulas and tree-like structures mentioned earlier
(see~\cite{Ershov1958,Goto1974}) indeed present with an ordering as soon as the
underlying tree representation is ordered.
This is the case for most trees used in computer science (\textit{e.g.}\
red-black trees, B-trees, etc.) and for all formulas involving non-commutative
operators.
The model we introduce thus allows for a more faithful modelling of a wide range
of objects.
We present here several results regarding DOAGs, as well as an extension of our
method to classical labelled DAGs.

As a first step of our analysis, we describe a recursive decomposition scheme
that allows us to study DOAGs using tools from enumerative combinatorics.
This allows us to obtain a recurrence formula for counting them, as well as a
polynomial-time uniform random sampler, based on the recursive method
from~\cite{NW1978}, giving full control over their number of vertices and edges.
Our decomposition is based on a ``vertex-by-vertex'' approach, that is we remove
one vertex at a time and we are able to describe exactly what amount of
information is necessary to reconstruct the graph.
This differs from the approach of Robinson to study DAGs, where all the sources
of a DAG are removed at once instead.
Although this is a minor difference, our approach allows us to easily account
for the number of edges of the graph, which is why our random sampler is able to
target DOAGs with a specific number of edges.
In terms of application, this means that we are able to efficiently sample
DOAGs of low density.
A second by-product of our approach is that it makes straightforward to bound
the out-degree of each vertex, thus allowing to sample DOAGs of low degree as
well.

In order to show the applicability of our method, we devise a similar
decomposition scheme for counting labelled DAGs with any number of vertices,
edges, and sources.
This allows us to transfer our results on DOAGs in the context of labelled DAGs.
Our new recurrence differs from the formula of Gessel~\cite{Gessel1996} in that
it does not resort to the inclusion-exclusion principle.
Our approach allows us to obtain an efficient uniform random sampler of labelled
DAGs with a prescribed number of vertices, edges, and sources.
Here again, in addition to giving control over the number of edges of the
produced objects, our approach can also be adapted to bound the out-degree of
their vertices.
To our knowledge, this is the first such sampler.

Finally, in a second part of our study of DOAGs, we focus on their asymptotic
behaviour and get a first result in this direction.
We consider the number~$D_n$ of DOAGs with~$n$ vertices, one source, and any
number of edges, and we manage to exhibit an asymptotic equivalent of an
uncommon kind:
\begin{equation*}
  D_n \sim c \cdot n^{-1/2} \cdot e^{n-1} \prod_{k=1}^{n-1} k!
  \quad\text{for some constant~$c > 0$.}
\end{equation*}
In the process of proving this equivalent, we state an upper bound on~$D_n$ by
exhibiting a super-set of the set of DOAGs of size~$n$, expressed in terms of
simple combinatorial objects: variations.
This upper-bound is close enough to~$D_n$ so that we can leverage it into an
efficient uniform rejection sampler of DOAGs with~$n$ vertices and any number of
edges.
Combined with an efficient anticipated rejection procedure, allowing to reject
invalid objects as soon as possible, this lead us to an asymptotically optimal
uniform sampler of DOAGs of size~$n$.

In terms of applications, our random generation algorithms enable
to experiment with the properties of the objects they model and with the average
complexity of algorithms operating on them.
A similar approach is for instance taken in~\cite{CP2024} where samplers for a
realistic class of Git graphs are developed in order to tackle the average
complexity of a new algorithm introduced in~\cite{Lecoq2024, CDL2024}.
Random testing a also an important application of random sampling, especially as
a building block for property-based testing, a now well-established framework
pioneered Claessen and Hughes in~\cite{CH2000}.

\bigskip

This paper extends an earlier article~\cite{GPV2021} with new results on the
asymptotics of DOAGs, with an optimal uniform random sampler for the case when
the number of edges is not prescribed, and covers a larger class of DOAGs and
DAGs by drooping a constraint on the number of sinks.
For the sake of completeness, the most important results and ideas
from~\cite{GPV2021} will be recalled in the present paper, but the reader will
have to refer the earlier article to get the full proof and algorithmic details.

\subsection{Outline of the paper}

In Section~\ref{sec:def}, we start by introducing the class of Directed Ordered
Acyclic Graphs and their recursive enumeration and describe a recursive
decomposition scheme allowing to count them.
In Section~\ref{sec:sampling:rec}, we quickly go over earlier results regarding
the random generation of DOAGs with a prescribed number of vertices, edges, and
sources.
The presentation given in this paper slightly generalises over the algorithm
given in~\cite{GPV2021} but the ideas and proofs remain unchanged.
Section~\ref{sec:dag} shows that our approach applies to labelled graphs as well
and opens the way for further research regarding this class.
We show that our method, when applied to labelled DAGs, yields a constructive
counting formula for them, that is amenable to efficient uniform random
generation with full control on the number of edges.
Then, in Section~\ref{sec:matrix}, we present a bijection between DOAGs and
class of integer matrices.
This bijection is the key result of this paper as it allows to understand the
structure of DOAGs in detail, and to obtain both asymptotic and algorithm
results in the following sections.
In Section~\ref{sec:asympt}, we present a first asymptotic result: we give an
asymptotic equivalent of the number of DOAGs of size~$n$ with any number of
sources and edges.
We also state some simple structural properties of those DOAGs.
In light of the matrix encoding and these asymptotic results, we design an
optimal uniform random sampler of DOAGs with a given number of vertices (but no
constraint on the number of edges), that is described in
Section~\ref{sec:sampling:rej}.

An implementation of all the algorithms presented in this paper is available
at~\url{https://github.com/Kerl13/randdag}.

\section{Definition and recursive decomposition}\label{sec:def}

In this section, we recall a model of directed acyclic graphs called ``Directed
Ordered Acyclic Graphs'' (or DOAGs) that we introduced in~\cite{GPV2021}. It is
similar to the classical model of unlabelled DAGs but where, in addition, we
have a total order on the outgoing edges of each vertex.
The presentation we opted for here slightly differs from that of~\cite{GPV2021}
but essentially defines the same objects, the only difference being that we now
allow several sinks for the sake of generality.

\begin{definition}[Directed Ordered Graph]\label{def:dog}
  A directed ordered graph (or DOG for short) is a triple~$(V, E,
  {(\prec_v)}_{v\,\in V \cup \{\top\}})$ where:
  \begin{itemize}
    \item $V$ is a finite set of vertices;
    \item $E \subset V \times V$ is a set of edges;
    \item for all~$v \in V$,~$\prec_v$ is a total order over the set of
      \emph{outgoing} edges of~$v$;
    \item and~$\prec_\top$ is a total order over the set of \emph{sources}
      of the graph, that is the vertices without any incoming edge.
  \end{itemize}
  Two such graphs are considered to be \emph{equal} if there exists a bijection
  between their respective sets of vertices that preserves both the edges and
  the order relations~$\prec_v$ and~$\prec_\top$.
\end{definition}

\begin{definition}[Directed Ordered Acyclic Graph]\label{def:doag}
  A directed ordered acyclic graph (or DOAG for short) is a directed ordered
  graph~$(V, E, {(\prec_v)}_{v\,\in V \cup \{\top\}})$ such that~$(V, E)$,
  seen as a directed graph, is acyclic.
\end{definition}

We study this class as a whole, however, some sub-classes are also of special
interest, in particular for the purpose of modelling compacted data structures.
Tree structures representing real data, such as XML documents for
instance~\cite{BLMN2015}, are rooted trees.
When these trees are compacted, the presence of a root translates into a unique
source in the resulting DOAG\@.
Similarly, DOAGs with a single sink will arise naturally when compacting trees
which bear a single type of leaves.
In particular the model of compacted binary trees, which can also be seen as a
class of cycle-free binary automata, has been shown have unusual combinatorial
properties in~\cite{EFW2020,EFW2021} and corresponds to a restriction of our
model with only binary nodes (and one sink).
For these reasons, we will also discuss how to approach the sub-classes of DOAGs
with a single source and/or a single sink in this document.

\begin{figure}[htb]
  \centering
  \includegraphics[scale=1]{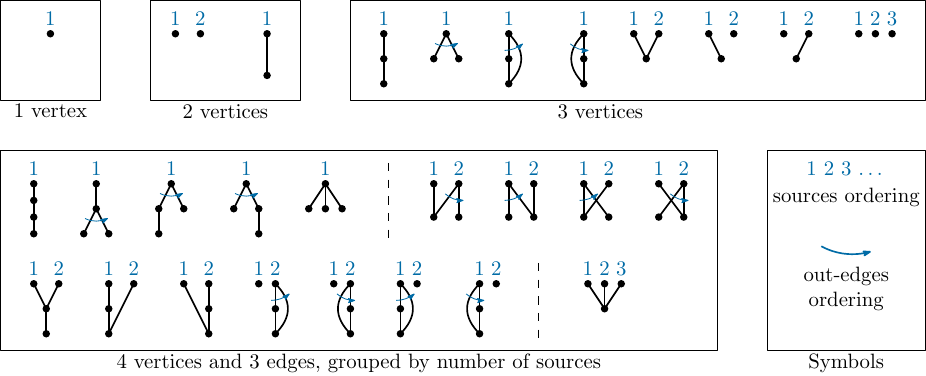}
  \caption{All DOAGs with respectly $1$ vertex, $2$ vertices, $3$ vertices,
    and simultaneously~$4$ vertices and~$3$ edges.
    All edges are implicitly oriented from top to bottom, the blue labels and
    arrows represent the sources and out-edges orderings (always from left to
    right).
  }\label{fig:firstdoags}
\end{figure}

In order to illustrate the definition, the first line of
Figure~\ref{fig:firstdoags} depicts all the DOAGs with at most~$3$ vertices and
the second line shows all the DOAGs with exactly~$4$ vertices and~$3$ edges.
There are 17 of them while there are~$95$ DOAGs with~$4$ vertices in total.

\subsection{Recursive decomposition}

We describe a canonical way to recursively decompose a DOAG into smaller
structures.
The idea is to remove vertices one by one in a deterministic order, starting
from the smallest source (with respect to their ordering~$\prec_\top$).
Formally, we define a decomposition step as a bijection between the set of DOAGs
with at least two vertices and the set of DOAGs given with some extra
information.

Let~$D$ be a DOAG with at least~$2$ vertices and consider the new graph~$D'$
obtained from~$D$ by removing its \emph{smallest} source~$v$ and its outgoing
edges.
We need to specify the ordering of the sources of~$D'$.
We consider the ordering where the \emph{new} sources of~$D'$ (those
that have been uncovered by removing~$v$) are considered to be in the \emph{same
order} (with respect to each other) as they appear as children of~$v$ and all
\emph{larger} than the other sources.
The additional information necessary to reconstruct~$D$ from~$D'$ is the
following:
\begin{enumerate}
  \item the number~$s \geq 0$ of sources of~$D'$ which have been uncovered by
    removing~$v$;
  \item the (possibly empty) set~$I$ of internal (non-sources) vertices of~$D'$
    such that there was an edge in~$D$ from~$v$ to them;
  \item the function $f:I \to \llbracket 1; s + |I| \rrbracket$ identifying the
    positions, in the list of outgoing edges of~$v$, of the edges pointing to
    an element of~$I$.
\end{enumerate}
More formally, this decomposition describes a bijection between DOAGs with at
least~$2$ vertices and quadruples of the form~$(D', s, I, f)$ where:
\begin{itemize}
  \item $D'$ is a DOAG (obtained by removing~$v$ from~$D$);
  \item $I$ is any subset of the internal vertices of~$D'$ (children of~$v$ in~$D$);
  \item $s$ is any integer between $0$ and the number of sources of~$D'$;
  \item and~$f: I \to \llbracket 1; s + |I| \rrbracket$ is an injective
    function (mapping the vertices of~$I$ to their positions in the list of
    children of~$v$ in~$D$).
\end{itemize}
In order to prove that this is indeed an bijection, we consider the inverse
transformation below.
Start with a quadruple~$(D', s, I, f)$ as described above.
Add a new source~$v$ in~$D'$ with~$s + |I|$ outgoing edges such that the~$i$-th
of these edges is connected to~$f^{-1}(i)$ when~$i \in f(I)$ and is connected to
one of the~$s$ largest sources of~$D'$ otherwise.
The~$s$ largest sources of~$D'$ must be connected to the new source exactly once
and in the same order as they appear in the list of sources of~$D'$.
The resulting graph is a DOAG and it is easy to check that this mapping and the
decomposition are inverses of each other.

\begin{figure}[htb]
  \centering
  \includegraphics[scale=.9]{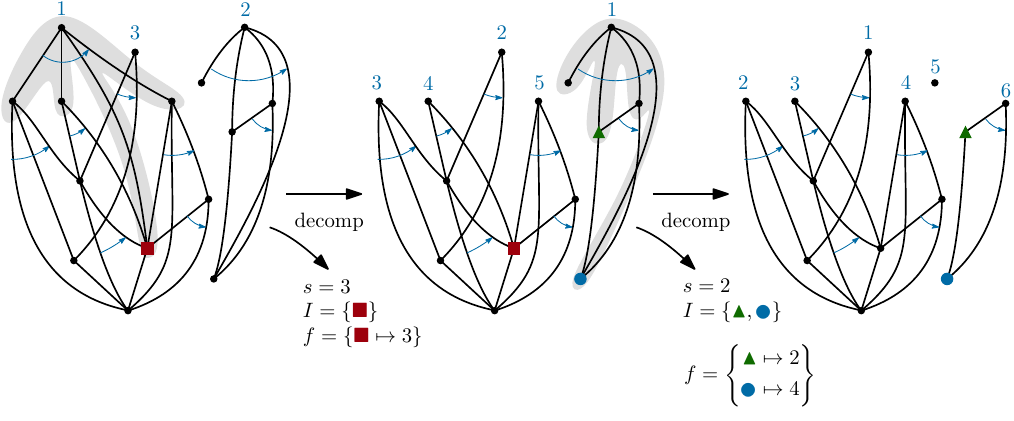}
  \caption{The two first steps of the recursive decomposition of a DOAG by
  removing sources one by one in a breadth first search (BFS) fashion.
  The edges are implicitly oriented from top to bottom and the order of the
  outgoing edges of each vertex is indicated by the thinner blue arrows (always
  from left to right here).
  The integer labels at each stage indicate the ordering of the sources.
  The big disk, square, and triangle are only here to highlight particular
  vertices involved with the functions~$f$ in the decomposition.}%
  \label{fig:decomposition}
\end{figure}

Note that the order in which the vertices are removed when iterating this
process corresponds to a variant of the BFS algorithm where only sources are
eligible to be picked next in the search, and their are picked in the order
described above.
Figure~\ref{fig:decomposition} illustrates this decomposition by applying the
first two steps on a large example DOAG\@.

This decomposition can be used to establish a recursive formula for counting
DOAGs, which is given below.
Let~$D_{n,m,k}$ denote the number of DOAGs with~$n$ vertices,~$m$ edges and~$k$
sources, then we have:
\begin{equation}\label{eq:recurrence}
  \begin{aligned}
    D_{1, m, k} &= \indicator{m=0 \, \land \, k=1} \\
    D_{n, m, k} &=
    \begin{cases}
      0 & \text{when}~k \leq 0 \\
      \sum_{p=0}^{n - k} \sum_{i=0}^p D_{n-1, m-p, k-1+p-i} \binom{n-k-p+i}{i}
      \binom{p}{i} i! & \text{otherwise,}
    \end{cases}
  \end{aligned}
\end{equation}
where~$p = s + i$ corresponds the out-degree of the smallest source.
The term~$\binom{n-k-p+i}{i} = \binom{n-k-s}{i}$ accounts for the choice of the
set~$I$ and the term~$\binom{p}{i} i{!}$ accounts for the number of injective
functions~$f: I \to \llbracket 1; p \rrbracket$.
The upper bound on~$p$ in the sum is justified by the fact that the out-degree
of any vertex can be at most the number of non-sources in the graph, that
is~$(n-k)$.

The decomposition scheme presented here differs from the approach described by
Robinson in~\cite{Robinson1970} as it operates on only one source at a time.
It is also reminiscent of the peeling processes used in map enumeration where
maps are decomposed one face at a time, see for instance~\cite{KMSW2019}.
However, the absence of ordering amongst the incoming edges of each vertex in
our setup renders those approaches inapplicable as is.

\subsubsection{Special sub-classes based on out-degree constraints}%
\label{sec:subclasses}

Since~$p = i + s$ is the out-degree of the removed source in the above
summation, it is easy to adapt this sequence for counting DOAGs with
constraints on the out-degree of the vertices.
For instance, DOAGs with only one sink are obtained by ensuring that every
vertex has out-degree at least one.
In other words, let the summation start at~$p=1$.
Note that restricting DOAGs to have only one single sink or one single source
ensures that they remain connected, however not all connected DOAGs are obtained
this way.
As another example, DOAGs with out-degree bounded by some constant~$d$ are
obtained by letting~$p$ range from~$0$ to~$\min(n-k, d)$.

The general principle is that the DOAGs whose vertices' out-degrees are
constrained to belong to a given set~$\mathcal P$, are enumerated by the
following generalised recurrence.
\begin{equation}\label{eq:recurrence:general}
  \begin{aligned}
    D^{\mathcal P}_{1, m, k} &= \indicator{m=0 \, \land \, k=1} \\
    D^{\mathcal P}_{n, m, k} &=
    \begin{cases}
      0 & \text{when}~k \leq 0 \\
      \sum_{p \in \mathcal P} \sum_{i=0}^p D^{\mathcal P}_{n-1, m-p, k-1+p-i}
      \binom{n-k-p+i}{i} \binom{p}{i} i! &  \text{otherwise,}
    \end{cases}
  \end{aligned}
\end{equation}

The first values of the sequence~$D_{n,m} = \sum_k D_{n,m,k}$ counting DOAGs by
number of vertices and edges only are given in Table~\ref{table:firstterms}.
Table~\ref{table:others} gives the first values of~$D^{\mathcal P}_n =
\sum_{m,k} D^{\mathcal P}_{n,m,k}$ for some relevant choices of~$\mathcal P$.
None of these sequences seem to appear in the online encyclopedia of integer
sequences (OEIS%
\footnote{\url{https://oeis.org/}}%
) yet.

\begin{table}[htb]
  \centering
  \caption{Number of DOAGs with~$n$ vertices and~$m$ edges for small values
  of~$n$ and~$m$.}\label{table:firstterms}

  \footnotesize
  \begin{tabular}{llp{37em}}
    \toprule
    $n$ & $D_n$ & $D_{n,m} = \sum_k D_{n,m,k}$ for~$m = 0, 1, 2, 3, \ldots$ \\
    \midrule
    $1$ & $1$ & $1$ \\
    $2$ & $2$ & $1, 1$ \\
    $3$ & $8$ & $1, 2, 3, 2$ \\
    $4$ & $95$ & $1, 3, 8, 17, 27, 27, 12$ \\
    $5$ & $4858$ & $1, 4, 15, 48, 139, 349, 718, 1136, 1272, 888, 288$ \\
    $6$ & $1336729$ & $1$, $5$, $24$, $100$, $391$, $1434$, $4868$, $14940$,
                      $40261$, $92493$, $175738$, $266898$, $310096$, $258120$,
                      $136800$, $34560$
  \end{tabular}
\end{table}

\begin{table}[htb]
  \centering
  \caption{Number of DOAGs with~$n$ vertices and a constrained set of allowed
  degrees.}\label{table:others}

  \footnotesize
  \begin{tabular}{p{.19\linewidth}p{.75\linewidth}}
    \toprule
    Restrictions & sequence \\
    \midrule
    $\mathcal P = \mathbb N$ \newline (all DOAGs) &
    $1$, $2$, $8$, $95$, $4858$, $1336729$, $2307648716$, $28633470321822$,
    $2891082832793961795$, $2658573971407114263085356$,
    $24663703371794815015576773905384$, $\ldots$ \\[.5em]
    $\mathcal P = \mathbb N$, $k=1$ \newline (1 source) &
    $1$, $1$, $4$, $57$, $3399$, $1026944$, $1875577035$, $24136664716539$,
    $2499751751065862022$, $2342183655157963146881571$,
    $22043872387559770578846044961204$, $\ldots$ \\[.5em]
    $\mathcal P = \mathbb N^\star$, $k=1$ \newline (1 source, 1 sink) &
    $1$, $1$, $3$, $37$, $2103$, $627460$, $1142948173$, $14701782996075$,
    $1522511169925136833$, $1426529804350999351686869$,
    $13426022673540053054145359653988$, $\ldots$ \\[.5em]
    $\mathcal P=\{0, 1, 2\}$, $k=1$ \newline (unary-binary) &
    $1$, $1$, $4$, $23$, $191$, $2106$, $29294$, $495475$, $9915483$,
    $229898277$, $6074257926$, $180460867600$, $5962588299084$, $\ldots$
  \end{tabular}
\end{table}

\section{Earlier results on counting and recursive sampling}\label{sec:sampling:rec}

In this section we summarise our earlier results on the counting and random
sampling problem for DOAGs when all three parameters (number of vertices, edges
and sources) are fixed.
The theorems are stated in a slightly more general setting here than
in~\cite{GPV2021} so as to capture all variants of the model as described in
Section~\ref{sec:subclasses}.
However, there is no technical difficulty in the generalisation so that the
proofs from~\cite{GPV2021} still apply, almost without modification.

We first present a utility result: computing all the numbers~$D_{n,m,k}^{\mathcal
P}$ up to a certain bound on~$n$,~$m$, and~$k$ can be done in polynomial time.
This is of moderate interest in itself, but this is a requirement for our
samplers, that compute theses values as a pre-processing step.
Our algorithm is based on the so-called ``recursive method'' from~\cite{NW1978}.

\subsection{Counting}

As mentioned above, tabulating the values of the sequence~$D_{n,m,k}^{\mathcal
P}$ can be done in polynomial time. This means that this counting pre-processing
step is tractable up to a certain point.

\begin{theorem}\label{thm:counting:cost}
  Let~$N, M > 0$ be two integers.
  And let~$\mathcal P$ be a subset of~$\mathbb N$ such that~$\mathcal P \cap
  \llbracket 0; n\rrbracket$ can be enumerated in linear time in~$n$.
  Computing~$D_{n, m, k}^{\mathcal P}$ for all~$n \leq N$, all~$m \leq M$, and
  all possible~$k$ can be done with~$O(N^4 M)$ multiplications of integers of
  size at most~$O(\max(M, N) \ln N)$.
\end{theorem}

The bound given here is independent of~$\mathcal P$ and thus pessimistic.
If~$\mathcal P$ is bounded (bounded out-degree DOAGs) or sparse, the algorithm
will perform better.
In practice, the cost of the counting process is actually the limiter factor for
the recursive sampler presented below.
Indeed, it is hard to reach sizes of the order of the thousands because of the
large amount of time and memory necessary to compute and store all the numbers.

\subsection{Recursive random sampling}

A straightforward application of the recursive method from Nijenhuis and Wilf
\cite{NW1978} leads to Algorithm~\ref{algo:sample:DOAG}, which is presented here
in a high level fashion.

\begin{algorithm}[htb]
  \caption{Recursive uniform sampler of DOAGs}%
  \label{algo:sample:DOAG}

  \begin{algorithmic}[1]
    \Require{Three integers~$(n, m, k)$ such that~$D_{n,m,k}^{\mathcal P} > 0$}
    \Ensure{A uniform random DOAG with~$n$ vertices (including $k$ sources)
    and~$m$ edges}
    \Function{UnifDOAG${}^{\mathcal P}$}{$n, m, k$}
      \If{$n=0$ \textbf{or} $n = 1$}
        {generate the (unique) DOAG with~$n$ vertex}
      \Else{}
        \State{\textbf{pick}~$(p, i)$ with probability
          $\displaystyle {D_{n-1, m-p, k-1+p-i}^{\mathcal P} \binom{n-k-p+i}{i} \binom{p}{i}
          i{!}} / {D_{n, m, k}^{\mathcal P}}$}%
          \label{algo:line:pick}
        \State{$D' \gets \text{\Call{UnifDOAG${}^{\mathcal P}$}{$n-1, m-p, k-1+p-i$}}$}%
        \label{algo:line:nsrc:begin}
        \State{$I \gets$ a uniform subset of size~$i$ of the inner vertices
        of~$D'$}\label{algo:line:subset}
        \State{$f \gets \text{a uniform injection from~$I$
        to~$\llbracket 1; p\rrbracket$}$}\label{algo:line:f}
        \return{$\decomp^{-1}(D', p-i, I, f)$}
      \EndIf{}
    \EndFunction{}
  \end{algorithmic}
\end{algorithm}

In~\cite[\S 3]{GPV2021}, we discuss how to implement
Algorithm~\ref{algo:sample:DOAG} efficiently.
In particular we suggest a data-structure to represent DOAGs that allows for
an efficient selection of the subset~$I$ at line~\ref{algo:line:subset} and the
function~$f$ at line~\ref{algo:line:f}.
In addition, implementation considerations are also given for the~\textbf{pick}
instruction at line~\ref{algo:line:pick}, which is the core of the ``recursive
method''.
As mentioned above, the numbers~$D_{n,m,k}^{\mathcal P}$ either have to be
pre-computed for Algorithm~\ref{algo:sample:DOAG} to work, or must be lazily
computed and memoised on the fly.

In practice, pre-computing all the necessary numbers to sample a uniform DOAG
with~$n = 50$ vertices (without any constraint on~$m$ and on the out-degree)
using our library already takes about~$8$ seconds on a standard laptop.
This running time rapidly increases, which makes the cost generating large
structures prohibitive.
However, when limiting the number of edges and using a finite set~$\mathcal P$,
one can achieve much better results.
For instance generating the four large bounded-degree DOAGs from
Figure~\ref{fig:bigdoag:random} takes about~$11$ seconds on the same laptop,
most of this time being spent in the pre-computation.

\begin{theorem}\label{thm:sample:cost}
  Algorithm~\ref{algo:sample:DOAG} computes a uniform random DOAG with~$n$
  vertices (among which~$k$ are sources) and~$m$ edges by
  performing~$O\left(\sum_{v} d_v^2 \right)$ multiplications of a small integer
  by a large integer, where~$v$ ranges over the vertices of the resulting graph
  and $d_v$ is the out-degree of~$v$.
\end{theorem}

Note that the sum~$\sum_v d_v^2$ is of the order of~$m^2$ in the worst case but
can be significantly smaller, in particular if~$\mathcal{P}$ is bounded or
sparse. In the best case we have~$d_v \sim \frac{m}{n}$ for most of the vertices
and as a consequence~$\sum_v d_v^2 \sim {m^2} / {n}$.
Also note that in order for the algorithm to be made generic in~$\mathcal P$, we
only have to use the sequence~$D_{n,m,k}^{\mathcal P}$ rather than~$D_{n,m,k}$,
which reflects the generality of the recursive method.

Four random DOAGs, drawn using Algorithm~\ref{algo:sample:DOAG} with~$\mathcal P
= \{0, 1, 2\}$,~$n=1250$,~$m=1300$ and~$k=1$ are pictured in
Figure~\ref{fig:bigdoag:random}.
As a comparison, a truncated version of the Git history of the linux kernel is
pictured on the left in Figure~\ref{fig:bigdoag:linux}.
Expectedly, the Git history looks more structured.
This is because developers work on short-lived branches, consisting of chains of
commits, generally starting from the main branch and merged back on main after
the new feature (or bug fix, etc.) has been completed, reviewed and accepted.

\begin{figure}[H]
  \centering
  \begin{subfigure}{.359\linewidth}
    \centering
    \includegraphics[width=.95\linewidth]{images-big-linux_70.pdf}
    \caption{The Git history of the Linux kernel, truncated at depth~$70$, as
      of the 14th of April 2025, just after the release of version
      \texttt{6.15-rc2}. It has~$1270$ vertices and~$1431$
      edges.\label{fig:bigdoag:linux}
  }
  \end{subfigure}
  \hfill
  \begin{subfigure}{.567\linewidth}
    \centering
    \noindent\includegraphics[width=.45\linewidth]{images-big-doag_1.pdf}
    \hfill
    \noindent\includegraphics[width=.45\linewidth]{images-big-doag_2.pdf}

    \noindent\includegraphics[width=.45\linewidth]{images-big-doag_3.pdf}
    \hfill
    \noindent\includegraphics[width=.45\linewidth]{images-big-doag_4.pdf}

    \caption{Four DOAGs drawn uniformly at random amongst all DOAGs with~$1250$
    vertices,~$1300$ edges and with out-degree bounded
    by~$2$.\label{fig:bigdoag:random}}
  \end{subfigure}
  \caption{The graphical representation of a (truncated) Git history and four
  bounded-degree random DOAGs.\label{fig:bigdoag}}
\end{figure}

This explains the chains of unary vertices and the triangular patterns.
Although the random DOAGs look more ``random'', they exhibit similar
smaller-scale structures such as triangular patterns and locally denser areas.
It has to be noted that in order to obtain those pictures, we had to reduce the
number of edges compared to the Git graph on the left as uniform DOAGs
with~$n=1270$ and~$m=1431$ are already visually too dense to look like a Git
graph.
We recall that the DOAG model presented in this paper aims at being a general
purpose modelling tool and thus does not integrate Git specific constraints.
In this regard, a random DOAG is not expected to have all the structural
properties of a Git graph.
However, the comparison in Figure~\ref{fig:bigdoag} showcases that the model can
be \emph{tweaked} (here by controlling~$m$ and the out-degree) in order to
resemble some application-inspired graphs.
In the case of Git, we obtain a somewhat similar shape.

\section{Extension to labelled DAGs}\label{sec:dag}

In this section we demonstrate how our decomposition scheme can be applied to
the classical model of labelled DAGs to obtain new recurrences on known
sequences.
In the 1990s, Gessel~\cite{Gessel1996} already obtained equations allowing to
count labelled DAGs by vertices, edges, and sources (and also sinks actually)
using a generating functions approach.
These equations involve the inclusion-exclusion principle which has one
drawback: they are usually not amenable to efficient random generation.
The reason for this is that subtractions translate into rejections in the
recursive algorithm, that are here too costly to be usable.
In the present paper, we derive new recurrences with a combinatorial meaning and
that do not involve the inclusion-exclusion principle.
As a consequence, we can obtain an efficient random sampler of DAGs with full
control over the number of vertices, sources, and edges.

As in Section~\ref{sec:def}, we present here a slight generalisation of the
formula given in~\cite[\S 4]{GPV2021} allowing to capture various classes of
labelled DAGs. We omit the proofs here as they follow a straightforward
adaptation of the arguments given in~\cite{GPV2021} and the recursive method.

\subsection{Recursive decomposition}

The key idea to our decomposition is to consider labelled DAGs with a
\emph{distinguished} source (this operation is called pointing) and to decompose
them by removing this source.
This describes a bijection between source-pointed labelled DAGs and labelled
DAGs endowed with some additional structure, in the same fashion as in
Section~\ref{sec:def} of the present paper.

Let~$\mathcal P$ be any subset of~$\mathbb N$ and let~$A_{n,m,k}^{\mathcal P}$
denote the number of labelled DAGs with~$m$ edges and $n$ vertices including~$k$
sources, and in which every vertex except the (first) sink has out-degree
in~$\mathcal P$.
The number of such DAGs with a distinguished (or pointed) source is given by~$k
\cdot A_{n,m,k}^{\mathcal P}$ since any of the~$k$ sources may be distinguished.
Let~$D$ denote one such DAG and let~$v$ denote its distinguished source.
Removing the distinguished source in~$D$ and decrementing the labels of the
vertices of higher label than~$v$ by one yields a regular vertex-labelled
DAG~$D'$ with~$n-1$ vertices.
Moreover, the three pieces of information that are necessary to reconstruct the
source are the following:
\begin{enumerate}
  \item the label~$\ell$ of the source~$v$ which has been removed;
  \item the set~$S$ of sources of~$D'$ which have been uncovered by
    removing~$v$;
  \item the set~$I$ of internal (non-sources) vertices of~$D'$ that were pointed
    at by~$v$.
\end{enumerate}
The reconstruction is then straightforward:
\begin{itemize}
  \item increment all the labels that are greater or equal to~$\ell$ by one;
  \item create a new vertex labelled~$\ell$ and ``mark'' it: this is the
    distinguished source;
  \item add edges from~$\ell$ to all the vertices from~$S$ and~$I$.
\end{itemize}
This decomposition is simpler than that of DOAGs because there is no ordering to
maintain here. Hence, any subset~$S$ of the set of sources of~$D'$ is licit
here.
The triplet~$(\ell, S, I)$ is thus not constrained which leads to the simple
counting formula, given below, where~$p$ denotes the out-degree of~$v$ (and thus
the cardinality of~$S \cup I$).
\begin{equation}\label{eq:recurrence:A}
  \begin{aligned}
    A_{1, m, k}^{\mathcal P} &= \indicator{m = 0~\wedge~k = 1} \\
    k A_{n, m, k}^{\mathcal P} &=
    \begin{cases}
      n
        \sum_{p \in \mathcal P \cap \llbracket 0; n - k\rrbracket} \sum_{i = 0}^p A_{n-1, m-p, k-1+p-i}^{\mathcal P}
        \binom{n-k-p+i}{i} \binom{k-1+p-i}{p-i}
        & \text{if~$1 \leq k$} \\
      0 & \text{otherwise.}
    \end{cases}
  \end{aligned}
\end{equation}
In the last equation:
\begin{itemize}
  \item the factor~$k$ on the left counts the number of ways to choose the
    distinguished source;
  \item the factor~$n$ on the right counts the number of ways to choose the
    label of the new source;
  \item and the two binomial coefficient count the number of ways to select the
    subsets~$I$ and~$S$.
\end{itemize}

When~$\mathcal P = \mathbb N$, we recover the sequence counting all labelled
DAGs, known as~\href{https://http://oeis.org/A003024}{\texttt{A003024}} in the
OEIS and first enumerated in~\cite{Robinson1970, Stanley1973, Robinson1973}.
For~$\mathcal P = \mathbb N^\star$ and with~$k=1$, we find the number of
labelled DAGs with a single source and a single sink, known to Gessel
in~\cite{Gessel1995,Gessel1996} and stored
at~\href{https://oeis.org/A165950}{\texttt{A165950}} in the OEIS.

\subsection{Random generation}

A recursive random sampling algorithm similar to
Algorithm~\ref{algo:sample:DOAG} from Section~\ref{sec:sampling:rec} can be
obtained from formula~\eqref{eq:recurrence:A}.
The only difference in methodology from  Algorithm~\ref{algo:sample:DOAG} is
that one has to deal with the marking of the sources here and thus the division
by~$k$ at the third line of~\eqref{eq:recurrence:A}.
It can be handled as follows: at every recursive call, first generate a labelled
DAG with a distinguished source (counted by~$k \cdot A_{n, m, k}$) and then
forget which source was distinguished.
Since the recursive formula for~$k \cdot A_{n, m, k}$ has no division, the
uniform sampler of marked DAGs is obtained using the standard recursive method.
Moreover, forgetting which source was marked does not introduce bias in the
distribution since all sources have the same probability to be marked.
A uniform random sampler of labelled DAGs with~$n$ vertices,~$k$ sources,
and~$m$ edges is described in Algorithm~\ref{algo:sample:dag}.

\begin{algorithm}[htb]
  \caption{Uniform random sampler of vertex-labelled DAGs.}%
  \label{algo:sample:dag}
  \begin{algorithmic}
    \Require{Three integers~$(n, m, k)$ such that~$A_{n,m,k}^{\mathcal P} > 0$}
    \Ensure{A uniform random labelled DAG with~$n$ vertices (including $k$
    sources and one sink),~$m$ edges, and in which every vertex (except the
    first sink) has out-degree in~$\mathcal P$.}
    \Function{unifDAG${}^{\mathcal P}$}{$n, m, k$}
      \If{$n = 0$ \textbf{or} $n=1$}
        generate the (unique) labelled DAG with~$n$ vertex
      \Else{}
        \State{\textbf{pick}~$(p, i)$ with probability
          $\displaystyle \frac{A_{n - 1, m - p, k-1+p-i}^{\mathcal P} \binom{n - k - p + i}{i}
          \binom{k - 1 + p-i}{p-i}}{A_{n, m, k}^{\mathcal P}}$}
        \assign{$D'$}{\Call{unifDAG${}^{\mathcal P}$}{$n - 1, m - p, k - 1 + p-i$}}
        \assign{$I$}{a uniform subset of size~$i$ of the inner vertices of~$D'$}
        \assign{$S$}{a uniform subset of size~$(p-i)$ the sources of~$D'$}
        \assign{$\ell$}{\Call{Unif}{$\llbracket 1; n \rrbracket$}}
        \State{relabel~$D'$ by adding one to all labels~$\ell' \geq \ell$}
        \return{the DAG obtained by adding a new source to~$D'$ with
        label~$\ell$ and with an outgoing edge to every vertex of~$I \cup S$}
      \EndIf{}
    \EndFunction{}
  \end{algorithmic}
\end{algorithm}

\section{Matrix encoding}\label{sec:matrix}

In this section, we introduce the notion of \emph{labelled transition matrices}
and give a bijection between DOAGs and these matrices, thus offering an
alternative point of view on DOAGs.
These results are key ingredients of the paper, since they enable us, in the
next two sections, to prove an asymptotic equivalence for the number of DOAGs
with~$n$ vertices, and to design a efficient uniform random sampler for those
DOAGs.
We also recall here the definition and basic properties of variations, which are
an elementary combinatorial object playing a central role in our analysis.

\subsection{The encoding}

The decomposition scheme described in Section~\ref{sec:def} corresponds to a
traversal of the DOAG\@.
This traversal induces a labelling of the vertices from~$1$ to~$n$, which allows
us to associate the vertices of the graph to these integers in a canonical way.
We then consider its transition matrix using these labels as indices.
Usually, the transition matrix of a directed graph~$D$ is defined as the
matrix~${(a_{i,j})}_{1 \leq i, j \leq n}$ such that~$a_{i,j}$ is~$1$ if there is
an edge from vertex~$i$ to vertex~$j$ in~$D$, and~$0$ otherwise.
This representation encodes the set of the edges of a DAG but not the edge
ordering of DOAGs.
In order to take this ordering into account, we use a slightly different
encoding.

\begin{definition}[Labelled transition matrix of a DOAG]
  Let~$D$ be a DOAG with~$n$ vertices.
  We associate the vertices of~$D$ to the integers from~$1$ to~$n$ corresponding
  to their order in the vertex-by-vertex decomposition.
  The \emph{labelled transition matrix} of~$D$ is the matrix~${(a_{i,j})}_{1
  \leq i, j \leq n}$ with integer coefficients such that~$a_{i,j} = k > 0$ if
  and only if there is an edge from vertex~$i$ to vertex~$j$ and this edge is
  the~$k$-th outgoing edge of~$i$.
  Otherwise~$a_{i,j} = 0$.
\end{definition}

\begin{figure}[htb]
  \centering
  \includegraphics[scale=1]{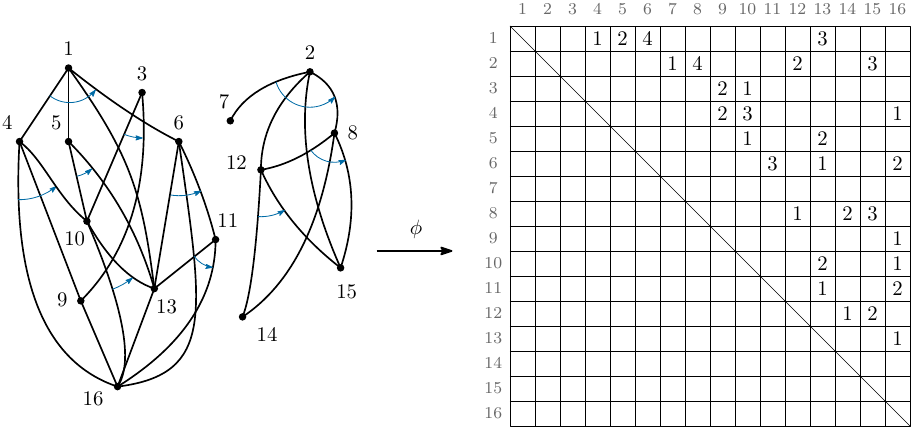}
  \caption{An example DOAG and its labelled transition matrix, the zeros are
  represented by the absence of a number.}%
  \label{fig:matrix}
\end{figure}

An example of a DOAG and its transition matrix are pictured in
Figure~\ref{fig:matrix}.
The thick lines are not part of the encoding and their meaning will be explained
later when we characterise which integer matrices can be a labelled transition
matrix.
Let~$\phi$ denote the function mapping a DOAG to its labelled transition matrix.
This function is clearly injective as the edges of the graph can be recovered as
the non-zero entries of the matrix, and the ordering of the outgoing edges of
each vertex is given by the values of the corresponding entries in each row.
Characterising the image of~$\phi$ however requires more work.

We can make some observations.
First, by definition of the traversal of the DOAG, the labelled transition
matrix of a DOAG is strictly upper triangular.
Indeed, since the decomposition algorithm removes one \emph{source} at a time,
the labelling it induces is a topological sorting of the graph.
Moreover, since the non-zero entries of row~$i$ encode the \emph{ordered} set of
outgoing edges of vertex~$i$, these non-zero entries form a permutation. More
formally:
\begin{itemize}
  \item a non-zero value cannot be repeated within a row;
  \item and if a row contains~$d \geq 1$ non-zero entries, then these are the
    integers from~$1$ to~$d$, in any order.
\end{itemize}
Informally, these two properties ensure that a matrix encodes a labelled DOAG (a
DOAG endowed with a labelling of its vertices) and that this labelling is a
topological sorting of the graph.
However, they are not enough to ensure that this topological sorting is
precisely the one that is induced by the decomposition.
The matrices satisfying these two properties will play an important role in the
rest of the paper.
We call them ``variation matrices''.

\begin{definition}[Variation]\label{def:var}
  A variation is a finite sequence of non-negative integers such that
  \begin{enumerate}
    \item each strictly positive number appears at most once;
    \item if~$0 < i < j$ and~$j$ appears in the sequence, then~$i$ appears too.
  \end{enumerate}
  The size of a variation is its length.
\end{definition}
For instance, the sequence~$(6, 2, 3, 0, 0, 1, 4, 0, 5)$ is a variation of
size~$9$ but the sequences $(1, 0, 3)$ and~$(1, 0, 3, 2, 3)$ are not variations
because the number~$2$ is missing in the first one and the second contains two
occurrences of the number~$3$.
Variations can also be defined as interleavings of a permutation with a
sequence of zeros.
One of the earliest references to these objects dates back to~1659 in
Izquierdo's \textit{Pharus Scientiarum}~\cite[Disputatio 29]{Izquierdo1659}.
They also appear in Stanley's book as the second entry of his \emph{Twelvefold
Way}~\cite[page 79]{Stanley2011}, a collection of twelve basic but fundamental
counting problems.
Knuth gives a few ancient references on this topic in~\cite{Knuth2011} and in an
quote (without reference) that can be found on the OEIS page of variations
at~\href{https://oeis.org/A007526}{A007526}.
Variations are relevant to our problem as they naturally appear as rows of the
labelled transition matrices defined in this section.
Some of their combinatorial properties will be exhibited in the next section.

\begin{definition}[Variation matrix]
  Let~$n > 0$ be a positive integer.
  A matrix of integers~${(a_{i,j})}_{1 \leq i, j \leq n}$ is said to be a
  variation matrix if
  \begin{itemize}
    \item it is strictly upper triangular;
    \item for all~$1 \leq i \leq n - 1$, the sub-row~${(a_{i,j})}_{i < j \leq
      n}$ is a variation (of size~$n-i$).
  \end{itemize}
  From an enumerative point of view, a variation matrix can be seen as a
  sequence of variations~$(v_1, v_2, \ldots, v_{n-1})$ where for all~$1 \leq i
  \leq n - 1$, the variation~$v_i$ has size~$i$.
\end{definition}

\begin{figure}[htb]
  \centering
  \includegraphics[scale=1]{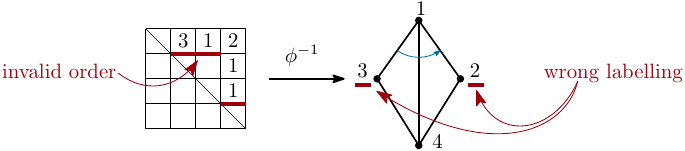}
  \caption{An example of a matrix of variations that cannot be obtained as a
  labelled transition matrix of a DOAG\@.
  The labelled DOAG that it encodes is not labelled according to the
  decomposition order.}%
  \label{fig:invalid}
\end{figure}

We have established that all labelled transition matrices of DOAGs are variation
matrices.
Note that the converse is not true.
For instance, the matrix pictured in Figure~\ref{fig:invalid} is a variation
matrix of size~$3$ that does not correspond to any DOAG\@.
The property of this matrix which explains why it cannot be the image of a DOAG
is pictured in red on the Figure.
The rest of this section is devoted to characterising which of those variation
matrices can be obtained as the labelled transition matrix of a DOAG\@.

\begin{figure}[htb]
  \centering
  \includegraphics[scale=1,page=2]{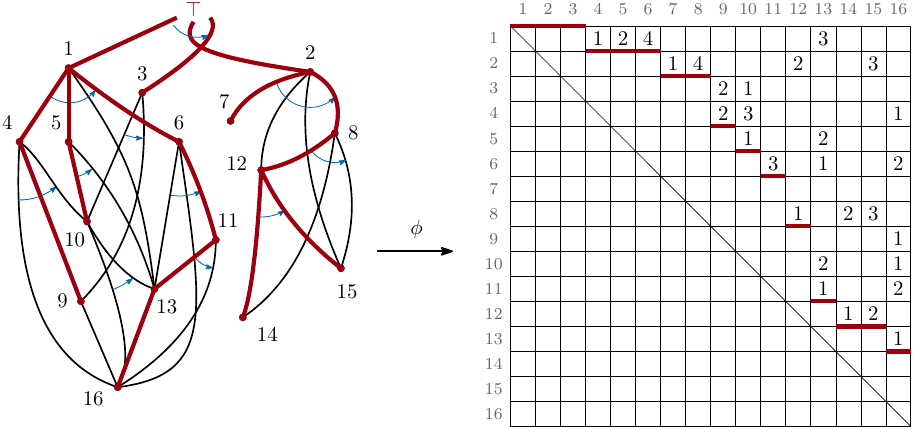}
  \caption{The same example as Figure~\ref{fig:matrix} with extra decorations to
    highlight the correspondence between the last incoming edge (in
    decomposition order) of each vertex and the last non-zero entry of the
    columns of its labelled transition matrix.
    An artificial~$\bot$ vertex, connected to every source, has been added in
    the graph in order to show that the thick red edges form a spanning tree of
    the graph.}
  \label{fig:matrix-tree}
\end{figure}

Consider a DOAG and its labelled transition matrix.
Note that in any column~$j$, the non-zero entry with the \emph{highest}
index~$i$ (that is in the lowest row on the picture with a non-zero element in
column~$j$) has a special role: it corresponds to the last edge pointing to
vertex~$j$ when decomposing the DOAG\@.
This is pictured in Figure~\ref{fig:matrix-tree} where we drew, the same DOAG as
in Figure~\ref{fig:matrix} and added:
\begin{itemize}
  \item on the right (in the matrix): thick red underlines to show the last
    non-zero entry of each column;
  \item on the left (in the graph): thick red decorations on the last incoming
    edge (in decomposition order) of each vertex.
\end{itemize}
When a column has no non-zero entry at all, the top line is pictured in thick
red instead.
This is the case in the three first columns of the matrix in
Figure~\ref{fig:matrix-tree}.
Still in the figure: in order to make those three extra lines correspond to
something in the graph, we added an artificial extra source, connected to all
other sources (there is a unique way to do this).
Those three extra edges are indeed the last incoming edges of the vertices
labelled 1, 2, and 3, that naturally correspond to the red part of the three
first columns of the matrix.
Note that the thick red edges in the graph on the left of
Figure~\ref{fig:matrix-tree} form a \emph{spanning tree} of the graph, and that
the labelling induced by the decomposition coincides exactly with the natural
BFS order of the tree.

Another important remark is that, when several underlined cells occur on the
same row~$i$ in the matrix, they correspond to several sources that are
discovered at the same decomposition step of the DOAG (upon removing the same
source).

\begin{wrapfigure}{r}{.2\linewidth}
  \centering
  \vspace*{-1pt}
  \includegraphics[width=.8\linewidth]{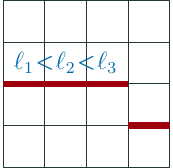}
  \caption{\small Same-row underlined entries are sorted.}%
  \label{fig:constraint:order}
  \vspace*{-2em}
\end{wrapfigure}
Recall that the decomposition algorithm sorts the labels of these new sources by
following the total order of the outgoing edges of vertex~$i$.
This implies that the underlined entries within the same row have to be
increasingly sorted (from left to right).
For instance, observe that there are three consecutive underlined cells in the
first row of the matrix in Figure~\ref{fig:matrix-tree}.
Indeed, when removing the first source of the DOAG on the left, we uncover three
new sources which are respectively in first, second and fourth position in the
outgoing edges order of the removed source.

\begin{wrapfigure}{r}{.2\linewidth}
  \centering
  \includegraphics[width=.9\linewidth]{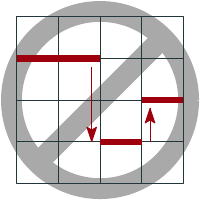}
  \caption{\small Thick red lines draw a descending staircase.}%
  \label{fig:constraint:dyck}
\end{wrapfigure}

\bigskip

A second key property is that if more than one underlined cell occur in a row of
the matrix, these are always the first non-zero entries in that row.
This is because, the decomposition algorithm consumes sources in the same order
in discovers them.
As a consequence, for a given vertex~$i$, those of its children that become
sources upon removing~$i$ will be processed before any other children, and thus
appear first in the list of the non-zero entries of the~$i$-th row.
Put differently, the red thick lines drawn in Figure~\ref{fig:matrix} is
visually a staircase that only goes down when moving toward the right of the
matrix.


The two properties that we just described actually characterise the variation
matrices that can be obtained as the labelled transition matrices of a DOAG\@.
This is stated in a more formal manner in Theorem~\ref{thm:matrix}.

\begin{theorem}\label{thm:matrix}
  All labelled transition matrices of DOAGs are variation matrices.
  Furthermore, let~$A = {(a_{i, j})}_{1 \leq i, j \leq n}$ be a variation
  matrix, and for all~$j \in \llbracket 1; n \rrbracket$, let~$b_j$ denote the
  largest~$i \leq n$ such that~$a_{i,j} > 0$ if such an index exists and~$0$
  otherwise.
  Then,~$A$ is the labelled transition matrix of some DOAG if and only if the
  two following properties hold:
  \begin{itemize}
    \item the sequence~$j \mapsto b_j$ is weakly increasing;
    \item whenever~$0 < b_j = b_{j+1}$, we have that~$a_{b_j,j} < a_{b_j,j+1}$.
  \end{itemize}
\end{theorem}

The sequence~$(b_j)_{1 \leq j \leq n}$ from the theorem is the formalisation of
the thick red lines from Figure~\ref{fig:matrix-tree}.
The first condition from the theorem corresponds to the ever-descending nature
of the ``staircase'', as illustrated in Figure~\ref{fig:constraint:dyck}
The second condition corresponds to the ordering of underlined cells within a
row, as illustrated in Figure~\ref{fig:constraint:order}.

\begin{proof}
  The fact that the labelled transition matrix of a DOAG is a variation matrix
  is clear from the definition.
  We prove the rest of the theorem in two steps.

  \medskip
  \noindent
  \textbf{Step 1: labelled transition matrices satisfy the conditions.}\quad
  Let~$D$ be a DOAG of size~$n$ and let~$A$ be its labelled transition matrix.
  Let~$b={(b_j)}_{1 \leq j \leq n}$ be defined as in the statement of the
  theorem.
  We shall prove that it satisfies the two properties of the theorem.
  The case~$n=1$ is trivially and we proceed by induction when~$n \geq 2$.

  When~$n \geq 2$, we can decompose~$D$ as a quadruplet~$(D', s, I, f)$ and we
  have that the labelled transition matrix of~$D'$ is the sub-matrix~$A'$ of~$A$
  obtained by removing its first row and first column, that
  is~$A'=(a_{i+1,j+1})_{1 \leq i, j \leq n - 1}$.
  We can define the sequence~$b' = (b_j')_{1 \leq j \leq n - 1}$ corresponding
  to~$A'$ similarly to the sequence~$b$.

  We distinguish between three cases.
  \begin{itemize}
    \item If~$j$ is such that~$b_j = 0$, then~$b_j \leq b_{j+1}$ automatically,
      and there is no second condition to check.

    \item If~$j$ is such that~$b_j = 1$, then~$b_{j+1}$ cannot be zero,
      otherwise that would mean that the vertex labelled~$(j+1)$ is a source
      of~$D$ but the vertex labelled~$j$ is not.
      Indeed, since the sources of~$D$ are processed before any other vertex by
      the decomposition algorithm, they get the smallest labels.
      Hence~$b_j \leq b_{j+1}$.

      In addition, if~$b_j = b_{j+1} = 1$, then the vertices labelled~$j$
      and~$(j+1)$ both become sources, \emph{at the same time}, upon removing
      the first vertex.
      By construction of the decomposition, they get labels in an order
      compatible with the order of the outgoing edges of the first source, and
      thus we have~$a_{1,j} < a_{1,j+1}$.

    \item Finally, if~$j$ is such that~$b_j \geq 2$, then we have~$b_j =
      b_{j-1}' + 1$.
      By induction we also have that~$b_j' \leq b_{j-1}'$, which, in particular,
      implies that there is at least one non-zero entry in the~$j$-th column
      of~$A'$ and thus in the~$(j+1)$-th column of~$A$.
      It follows that~$b_{j+1} = b_j' + 1$ and finally~$b_j \leq b_{j+1}$
      and~$b_j = b_{j+1} \implies a_{b_j,j} < a_{b_j,j+1}$ by induction.
  \end{itemize}

  \medskip
  \noindent
  \textbf{Step 2: any matrix satisfying the conditions is a labelled transition
  matrix.}\quad
  Let~$A$ be a variation matrix of size~$n$ and let~$b$ be as in the statement
  of the theorem and satisfying the two given properties.
  We shall prove that~$A$ is the image by~$\phi$ of some DOAG\@.

  Let~$V = \llbracket 1; n \rrbracket$ and~$E = \left\{(i, j) \in \llbracket 1;
  n \rrbracket^2~|~a_{i,j} > 0\right\}$.
  We have that~$(V, E)$ defines an acyclic graph since~$A$ is strictly
  upper-triangular.
  In addition, for each~$v \in V$, define~$\prec_v$ to be the total order on the
  outgoing edges of~$v$ in~$(V, E)$ such that~$u \prec_v u'$ if and only
  if~$a_{v,u} < a_{v,u'}$ in~$A$.
  This is well defined since the outgoing edges of~$v$ are precisely the
  integers~$j$ such that~$a_{v,j} > 0$ and since the non-zero entries of the
  row~$v$ are all different by definition of variation matrices.
  Finally, define~$\prec_\top$ to be the total order on the sources of~$(V, E)$
  such that~$u \prec_\top v$ if and only if~$u < v$ as integers.
  Let~$D$ be the DOAG given by~$(V, E, {(\prec_v)}_{v\,\in V \cup
  \{\top\}})$.

  Remember that DOAGs are considered up to a permutation of their vertices that
  preserves~$E$ and~$\prec$.
  In order to finish this proof, we have to check that the particular labelling
  encoded by~$V$ is indeed the labelling induced by the decomposition of~$D$.
  Then it will be clear that~$\phi(D) = A$ and we will thus have exhibited a
  pre-image of~$A$.

  First, since~$A$ is strictly upper-triangular, its first column contains only
  zeros and thus~$1$ is necessarily a source of~$D$.
  In addition, by definition of~$\prec_\top$, it must be the smallest
  source.
      Then, upon removing~$i$, one of two things can happen:
      \begin{itemize}
        \item either~$D$ has more than one source, in which case~$2$ is the
          second source by monotony of the sequence~${(b_j)}_{1 \leq j \leq n}$;
        \item or~$1$ was the unique source of~$D$, in which case the next source
          to be processed is its first child.
          The children of~$1$ are the integers~$j$ such that~$b_j = 1$.
          By monotony of~$b_j$ again (or triangularity of the matrix),~$2$ is
          necessarily a child of~$1$.
          Moreover, by the second property of the sequence~$b$, we have that for
          all~$j < j'$ such that~$b_j = b_{j'} = 1$,~$a_{1,j} < a_{1,j'}$.
      \end{itemize}
      In both case, we proved that~$2$ is the second vertex to be processed.
      We can then repeat this argument on the DOAG obtained by removing~$1$,
      which corresponds to the matrix~${(a_{i,j})}_{2 \leq i, j \leq n}$ and
      conclude by induction.\qedhere
\end{proof}

We have now established that the encoding~$\phi$ of DOAGs as labelled transition
matrices is a bijection from DOAGs to the matrices described in
Theorem~\ref{thm:matrix}.
From now on, we will write ``a labelled transition matrix'' to refer to such a
matrix.
We can also state a few simple properties of these matrices.
By definition we have that
\begin{itemize}
  \item the number of vertices of a DOAG is the dimension of its labelled
    transition matrix;
  \item the number of edges of a DOAG is the number of non-zero entries of the
    matrix;
  \item the sinks of the DOAG correspond to the zero-filled rows of the matrix;
  \item the sources of the DOAG correspond to the zero-filled columns of the
    matrix.
\end{itemize}
Furthermore, the first property of the sequence~${(b_j)}_{1 \leq j \leq n}$
defined in Theorem~\ref{thm:matrix} implies that the zero-filled columns of the
matrix must be contiguous and on the left of the matrix.
The number of sources of the DOAG is thus the maximum~$j$ such that column~$j$
is filled with zeros.

We will see in the next section that working at the level of the labelled
transition matrices, rather than at the level of the graphs, is more handy to
exhibit asymptotic behaviours.
This will also inspire an efficient uniform random sampler of DOAGs with~$n$
vertices in Section~\ref{sec:sampling:rej}.

\section{Asymptotic results}\label{sec:asympt}

The characterisation of the labelled transition matrices of DOAGs gives a more
\emph{global} point of view on them compared to the decomposition given earlier,
which only looks \emph{locally} around one source.
By approaching the counting problem from the point of matrices, we manage to
provide lower and upper bounds on the number of DOAGs with~$n$ vertices (and any
number of edges).
These bounds are precise enough to give a good intuition on the asymptotic
behaviour of these objects, and we then manage to refine them into an asymptotic
equivalent for their cardinality.
Building on this same approach, we provide in Section~\ref{sec:sampling:rej} an
efficient uniform sampler of DOAGs with~$n$ vertices.

This section is mostly devoted to proving Theorem~\ref{thm:Dnequiv}.
Sub-sections~\ref{sec:fstbounds} to~\ref{sec:cstapprox} present the general
approach and give all the intermediate results that are necessary to prove
Theorem~\ref{thm:Dnequiv}.
We conclude this section by giving asymptotic estimations of two relevant
parameters of DOAGs under the uniform model: their number of sources and edges.
We obtain these two last results by leveraging the work done in this section on
the matrix point of view on DOAGs.

\begin{theorem}\label{thm:Dnequiv}
  There exists a constant~$c > 0$ such that the number~$D_n$ of DOAGs with~$n$
  vertices and the number~$D_n^\star$ of such DOAGs having only one source
  satisfy
  \begin{align*}
    D_n
    &= \left(1 + \bigO{n^{-1}}\right) D_n^\star
    = \frac{c}{\sqrt{n}} \ e^{n-1} \ \sfact{(n-1)} \left(1 + \bigO{n^{-1}}\right)
    \\
    &\sim c \cdot e^{\zeta'(-1) - 1} \cdot n^{-7 / 12} \cdot
    {\big(e\sqrt{2 \pi}\big)}^n \! \cdot
    e^{- \frac{3}{4} n^2} \cdot
    n^{n^2 / 2}
  \end{align*}
  where~$\sfact{k} = \prod_{i=0}^k {i!}$ denotes the \emph{super factorial}
  of~$k$.
\end{theorem}

The super factorial provides a concise way to express this equivalent and also
reflects the relation between DOAGs and variation matrices, which will be
further developed in this section.

\subsection{First bounds on the number of DOAGs
with~\texorpdfstring{$\boldsymbol n$}{n} vertices}\label{sec:fstbounds}

Let~$D_n = \sum_{m,k} D_{n,m,k}$ denote the number of DOAGs with~$n$ vertices
and any number of sources and edges.
By Theorem~\ref{thm:matrix}, all labelled transition matrices are variation
matrices.
A straightforward upper bound for~$D_n$ is thus given by the number of variation
matrices of size~$n$.

\begin{lemma}[Upper bound on the number of DOAGs]%
  \label{lem:doag:upper}
  For all~$n \geq 1$, the number~$D_n$ of DOAGs of size~$n$ satisfies
  \begin{equation*}
    D_n \leq \sfact{(n-1)} e^{n-1}
  \end{equation*}
  where~$\sfact{k} = \prod_{i=0}^k {i!}$ denotes the \emph{super factorial}
  of~$k$.
\end{lemma}
The term ``super factorial'' seems to have been coined by Sloane and Plouffe
in~\cite[page~228]{SP1995} but this sequence had been studied before that,
in~1900, by Barnes~\cite{Barnes1900} as the integer values of
the ``G-function''.
In fact, if~$G(z)$ denotes the complex-valued G-function of Barnes, we have the
identity~$G(n+2) = \sfact{n}$ for all integer~$n$.
Barnes also gives the following equivalent.

\begin{lemma}[Asymptotic estimation of the super-factorial~\cite{Barnes1900}]%
\label{lem:barnes}
  When~$n \to \infty$, we have
  \begin{equation*}
    \sfact{(n-1)} = G(n+1) \sim
      e^{\zeta'(-1)} \cdot
      n^{-1 / 12} \cdot
      {\big(\sqrt{2 \pi}\big)}^n \cdot
      e^{- \frac{3}{4} n^2} \cdot
      n^{n^2 / 2}
  \end{equation*}
  where~$\zeta$ denotes the Riemann zeta function.
\end{lemma}

In order to prove Lemma~\ref{lem:doag:upper}, we first need to give estimates
for the number~$v_n$ of variations of size~$n$.

\begin{lemma}[Exact and asymptotic number of variations]\label{lem:var}
  For all~$0 \leq p \leq n$, the number~$v_n$ of variations of size~$n$, and the
  number~$v_{n,p}$ of variations of size~$n$ containing exactly~$p$ zeros, are
  respectively given by
  \begin{equation*}
    v_n = n! \sum_{j = 0}^n \frac{1}{j!}
    \quad\text{and}\quad
    v_{n,p} = \frac{n!}{p!}
  \end{equation*}
  As a consequence~$v_n \leq e\cdot n{!}$ and~$v_n = e\cdot n! + o(1)$.
\end{lemma}
\begin{proof}
  Let~$0 \leq p \leq n$, a variation of size~$n$ containing exactly~$p$ zeros is
  the interleaving of a permutation of size~$(n-p)$ with an array of zeros of
  size~$p$.
  As a consequence
  \begin{equation*}
    v_{n,p} = \binom{n}{p} (n-p)! = \frac{n!}{p!}.
  \end{equation*}
  We then get~$v_n$ and the asymptotic estimate by summation:
  \begin{equation*}
    v_n
    = \sum_{p=0}^n v_{n,p}
    = n! \sum_{p=0}^n \frac{1}{p!}
    = n! \left(\sum_{p=0}^\infty \frac{1}{p!} - \sum_{p=n+1}^\infty
    \frac{1}{p!}\right)
    = e n! - \sum_{p=n+1}^\infty \frac{n!}{p!},
  \end{equation*}
  which allows to conclude since the last sum is~$\sum_{p>n} \frac{n!}{p!} =
  O(n^{-1})$.
\end{proof}

The proof of Lemma~\ref{lem:doag:upper} follows from this lemma.
\begin{proof}[Proof of Lemma~\ref{lem:doag:upper}]
  By inclusion, there are less DOAGs of size~$n$ that there are variation
  matrices.
  In addition, a variation matrix is given by a sequence~$v_1, v_2, \ldots,
  v_{n-1}$ of variations such that for all~$i$,~$v_i$ is of size~$i$.
  We thus have the following upper bound for~$D_n$:
  \begin{equation*}
    D_n
    \leq \prod_{i = 1}^{n - 1} v_{n-i}
    \leq \prod_{i = 1}^{n - 1} e \cdot (n-i)!
    = \sfact{(n - 1)} e^{n - 1}.
    \qedhere
  \end{equation*}
\end{proof}

\begin{wrapfigure}{r}{.2\linewidth}
  \centering
  \includegraphics[width=.9\linewidth]{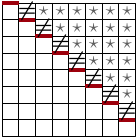}
  \caption{Lower-bound on the set of labelled transition matrices.}%
  \label{fig:proof:lower}
\end{wrapfigure}
Obtaining a lower bound on~$D_n$ requires to find a subset of the possible
labelled transition matrices described in Theorem~\ref{thm:matrix} that is both
easy to count and large enough to capture a large proportion of the DOAGs.
One possible such set is that of the labelled transition matrices which have
no zero values on the super-diagonal~${(a_{i, i+1})}_{1 \leq i < n}$.
These matrices are picture in Figure~\ref{fig:proof:lower} on the right.

These correspond to DOAGs such that, at every step of the decomposition, we
have only one source and thus uncover exactly one new source.
In such matrices, the properties of the sequence~${(b_j)}_{1 \leq j \leq n}$
from Theorem~\ref{thm:matrix} are automatically satisfied.
Intuitively, forcing the super-diagonal to be positive still leaves a
large amount of free space on the right of that diagonal to encode many possible
DOAGs, so it should be expected that it gives a decent lower bound.

\begin{lemma}[A first lower bound on the number of DOAGs]%
  \label{lem:doag:firstlower}
  There exists a constant~$A > 0$ such that for all~$n \geq 1$, we have
  \begin{equation*}
    \frac{A}n \sfact{(n - 1)} e^{n - 1} \leq D_n.
  \end{equation*}
\end{lemma}

\begin{proof}
  In a labelled transition matrix with positive values on the super-diagonal,
  the~$i$-th row can be seen as a variation of size~$(n-i)$ that does not start
  with a zero.
  Moreover, the number of variations of size~$n$ starting with a zero is
  actually the number of variations of size~$(n-1)$ so that the number of
  possibilities for the~$i$-th row of the matrix we count here is~$(v_{n-i} -
  v_{n-i-1})$.
  In addition, by Lemma~\ref{lem:var}, we also have that
  \begin{equation}
    v_n - v_{n-1}
    = e \cdot n! - e \cdot (n-1)! + o(1)
    = e \cdot n! \cdot \frac{n-1}{n} \left(1 +
    O\left(\frac{1}{n!}\right)\right).
  \end{equation}
  Note that when~$i = n-1$, we have~$v_{n-i} - v_{n-i-1} = v_1 - v_0 = 1$.
  Indeed, the row of index~$i = n-1$ contains only the number~$1$ in the super
  diagonal since at the last step of the decomposition, we have two connected
  vertices and there is only one such DOAG\@.
  Setting aside this special case, which does not contribute to the product, we
  get the following lower bound for~$D_n$:
  \begin{equation}
    D_n
    \geq \prod_{i=1}^{n-2} (v_{n-i} - v_{n-i-1})
    = e^{n-2} \sfact{(n-1)} \prod_{i = 1}^{n - 2} \frac{n - 1 - i}{n - i} \prod_{i
    = 1}^{n - 1} \left(1 + O\left(\frac{1}{(n - i)!}\right)\right)
  \end{equation}
  where the first product telescopes and yields~$\frac{1}{n-1}$ and the second
  one converges to a constant as~$n \to \infty$.
  This allows to conclude the proof the lemma.
\end{proof}

Although they are not precise enough to obtain an asymptotic equivalent for the
sequence~$D_n$, these two bounds already give us a good understanding of the
behaviour of~$D_n$.
First of all, they let appear a ``dominant'' term of the form~$\sfact{(n-1)}
e^{n-1}$, which is uncommon in combinatorial enumeration.
And second, it tells us we only make a relative error of the order of~$O(n)$
when approximating~$D_n$ by~$\sfact{(n-1)} \cdot e^{n-1}$.
We will prove an asymptotic equivalent for the remaining polynomial term, but in
order to obtain this, we first need to slightly refine our lower bound so that
the ``interval'' between our two bounds is a little narrower than~$\bigO{n}$.

\begin{lemma}[A better lower bound for the number of DOAGs]%
  \label{lem:doag:lower}
  There exists a constant~$B > 0$ such that, for all~$n \geq 1$, we have
  \begin{equation*}
    D_n \geq B \frac{\ln(n)}{n} \sfact{(n - 1)} e^{n - 1}.
  \end{equation*}
\end{lemma}

\begin{proof}
  In order to obtain this lower bound, we count the number of valid labelled
  transition matrices such that \emph{all but exactly one} of the cells on the
  super-diagonal have non-zero values.
  Furthermore, in order to avoid having to deal with border cases, we assume
  that the unique zero value on the super-diagonal appears between~$i=2$
  and~$i=n-2$.
  Let~$2 \leq i \leq n - 2$ and let us consider those matrices~${(a_{p,q})}_{1
  \leq p, q \leq n}$ such that~$a_{i, i+1} = 0$.
  Those matrices are illustrated in Figure~\ref{fig:proof:betterlower}

  \begin{figure}[htb]
    \centering
    \includegraphics[scale=1]{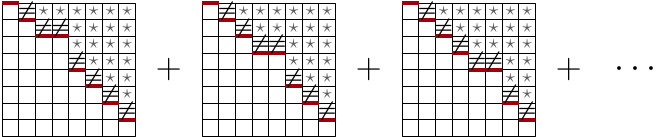}
    \caption{Illustration of the matrices with only one zero on the super-diagonal}
    \label{fig:proof:betterlower}
  \end{figure}

  The differences between these matrices and those enumerated in the proof of
  the previous lemma are the following (assuming~$i$ is that unique index such
  that~$a_{i,i+1}=0$).
  \begin{enumerate}
    \item On row~$i - 1$, the two first cells on the right of the
      diagonal~($a_{i-1, i}$ and~$a_{i-1,i+1}$) must have positive values and
      must be in increasing order.
      In the case of~$a_{i-1,i}$, this is because it is on the super diagonal.
      And for~$a_{i-1,i+1}$, this is because~$a_{i,i+1} = 0$: since it is above
      this cell, and since the cell~$a_{i-1,1}$ on its left is non-zero, it must
      be non-zero.
      Otherwise the condition from Theorem~\ref{thm:matrix} are
      violated.
      In terms of DOAGs, this means that the vertex~$i-1$ produces two new
      sources when it is removed but vertex~$i$ produces none.\label{item:iminus1}
    \item On row~$i$, any variations of size~$(n-i)$ starting by a zero is
      allowed.\label{item:i}
  \end{enumerate}
  We get the number of variations of size~$n$ starting by two increasing
  positive values (condition~\ref{item:iminus1} above) by inclusion-exclusion.
  That is,
  \begin{itemize}
    \item consider all the variations of size~$n$~($v_n$ possibilities);
    \item remove the number of variations that have a zero in first
      position~($v_{n-1}$ possibilities);
    \item remove the number of variations that have a zero in second
      position~($v_{n-1}$ possibilities);
    \item add the number of variations that start with two zeros, because they
      have been removed twice in the two previous lines~($v_{n-2}$
      possibilities);
    \item and finally, divide by two because only half of these matrices have
      their first values in increasing order.
  \end{itemize}
  This yields the following formula for counting such variations:
  \begin{equation*}
    \frac{v_n - 2 v_{n-1} + v_{n-2}}{2}
    \underset{n\to\infty}\sim \frac{e}{2} \cdot n!.
  \end{equation*}
  As a consequence, the total number of labelled transition matrices considered
  at the beginning of the proof, such that~$a_{i,i+1} = 0$, is given by
  \begin{align*}
    & \frac{v_{n-i-1} - 2 v_{n-2-i} + v_{n-3-i}}{2}
    \cdot v_{n-i-1}
    \cdot \prod_{\substack{1 \leq p \leq n-1 \\ p \not\in \{i-1, i\}}} (v_{n-p} - v_{n-p-1})
    \\
    &= \frac{(v_{n-i-1} - 2 v_{n-2-i} + v_{n-3-i}) v_{n-i-1}}%
    {2(v_{n-i+1} - v_{n-i})(v_{n-i} - v_{n-i-1})}
    \cdot \prod_{p=1}^{n-1} (v_{n-p} - v_{n-p-1}).
  \end{align*}
  By summing over~$2 \leq i \leq n - 2$, we get
  \begin{equation}\label{eq:proof:lb}
    \sum_{i=2}^{n-2}
    \frac{(v_{n-i-1} - 2 v_{n-2-i} + v_{n-3-i}) v_{n-i-1}}%
    {2(v_{n-i+1} - v_{n-i})(v_{n-i} - v_{n-i-1})}\cdot
    \prod_{p=1}^{n-1} (v_{n-p} - v_{n-p-1}).
  \end{equation}
  The fraction in the last equation is equivalent to~$\frac{1}{2(n-i)}$
  when~$n-i \to \infty$.
  So after a change of variable, the above sum is equivalent to
  \begin{equation*}
    \sum_{i=2}^{n-2}
    \frac{(v_{i-1} - 2 v_{i-2} + v_{i-3}) v_{i-1}}%
    {2(v_{i+1} - v_i)(v_{i} - v_{i-1})}
    \sim\sum_{i=2}^{n-2} \frac{1}{2i}
    \sim \frac{\ln(n)}{2}
  \end{equation*}
  In addition, we know from the proof of Lemma~\ref{lem:doag:firstlower} that
  the product in equation~\eqref{eq:proof:lb} is equivalent to~$\frac{c}{n}
  e^{n-1} \sfact{(n-1)}$ for some constant~$c$, which allows to conclude.
\end{proof}

\subsection{Obtaining the polynomial term by bootstrapping}

Let us denote~$P_n$ the polynomial term in~$D_n$, that is the quantity
\begin{equation*}
  P_n \overset{\text{def}}= \frac{D_n}{\sfact{(n-1)} e^{n-1}}.
\end{equation*}
We have proved above that for some constant~$B > 0$, we have~$B \frac{\ln(n)}{n}
\leq P_n \leq 1$.
A consequence of these inequalities is that for all~$k \in \ZZ$, we have
\begin{equation}\label{eq:Pnplusk}
  P_{n+k} \leq 1 \leq \frac{n}{B \ln(n)} P_n \underset{n \to \infty}= o(n P_n).
\end{equation}
Note that we did the extra work in Lemma~\ref{lem:doag:lower} in order to get
the extra~$\ln(n)$ factor that is crucial to get the~$o$ term.
Equation~\eqref{eq:Pnplusk} allows to justify that~$P_{n+k} / n$ is negligible
compared to~$P_n$, for any constant values of~$k$.
Although intuitively the contrary would be surprising, this fact is not clear
\textit{a priori} as an arbitrary polynomial sequence~$P_n$ could have violent
oscillations for some values of~$n$.
This is a key ingredient for proving an asymptotic equivalent for~$P_n$.

\begin{figure}[htb]
  \centering
  \includegraphics[scale=1]{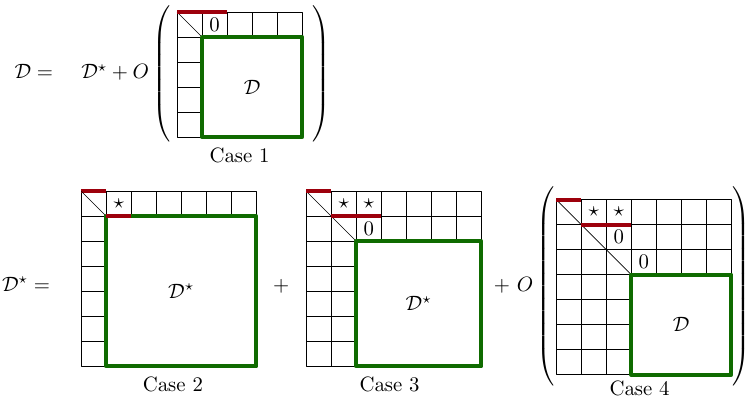}
  \caption{Decomposition of DOAG labelled transition matrices based or their
  content near the top of the diagonal.
  The symbols~$\mathcal D$ and~${\mathcal D}^\star$ respectively represent the
  set of all possible DOAG labelled transition matrices the set of all of those
  matrices such that~$a_{1,2} > 0$.
  The stars~$(\star)$ represent strictly positive values.}%
  \label{fig:mat:decomp}
\end{figure}

To refine our knowledge on the sequence~$P_n$, we use a decomposition of the
labelled transition matrices focused on the values it takes near the diagonal on
its first rows.
We categorise the possible labelled transition matrices~${(a_{i,j})}_{1 \leq i,
j \leq n}$ into the four following cases.
\begin{description}
  \item[Case 1:~$\boldsymbol{a_{1,2} = 0}$.] In this case, the first source is
    not connected to the second vertex and the matrix has thus more than one
    source.
    The first row of such a matrix is a variation of size~$(n-2)$ and the lower
    part~${(a_{i,j})}_{2 \leq i, j \leq n}$ encodes a DOAG of size~$(n-1)$, the
    DOAG obtained by removing the first source.
    However, it is important to note that not all combinations of a size-$(n-2)$
    variation and a size-$(n-1)$ matrix yield a valid size-$n$ labelled
    transition matrix.
    For instance, a variation of the form~$v = (0, 1, 0, 2, \ldots)$ and a lower
    matrix with at least three sources cannot be obtained together as they would
    violate the constraints of Theorem~\ref{thm:matrix}.
  \item[Case~2:~$\boldsymbol{a_{1,2} > 0 \wedge a_{2, 3} > 0}$.] In this case,
    the first row is a variation of size~$(n-1)$ starting by a positive value,
    and the lower part~${(a_{i,j})}_{2 \leq i, j \leq n}$ encodes a DOAG of
    size~$(n-1)$ with exactly one source, again obtained by removing the first
    source.
    This second fact is a direct consequence of~$a_{2,3} > 0$.
    Here, all such pairs can be obtained.
  \item[Case~3:~$\boldsymbol{a_{1,2} > 0 \wedge a_{2,3} = 0 \wedge a_{3,4} >
    0}$.] In this case the lower part~${(a_{i,j})}_{3 \leq i, j \leq n}$ encodes
    a DOAG of size~$n-2$ with exactly one source, the first row is necessarily a
    variation of size~$(n-1)$, starting with two positive increasing values, and
    the second row is a variation of size~$(n-2)$ starting by a zero.
    Here again this decomposition is exact: all such triplets can be obtained
    here.
  \item[Case~4:~$\boldsymbol{a_{1,2} > 0 \wedge a_{2, 3} = a_{3, 4} = 0}$.]
    Finally, this case captures all the remaining matrices.
    The first row is a variation of size~$(n-1)$, the second and third rows are
    variations of sizes~$(n-2)$ and~$(n-3)$ starting with a zero, and the lower
    part~${(a_{i,j})}_{4 \leq i, j \leq n}$ encodes a size-$(n-3)$ DOAG\@.
    Of course, not all such quadruples can be obtained, but this
    over-approximation will be enough for our proof.
\end{description}
This decomposition into four different cases is illustrated in
Figure~\ref{fig:mat:decomp} where~$\mathcal D$ represents the set of all
possible DOAG labelled transition matrices and~${\mathcal D}^\star$ represents
all of those matrices that encode a single-source DOAG\@.

We compute the contributions to~$D_n$ coming from each of these four terms
described above.
Let us denote by~$D_n^\star$ the number of DOAG of size~$n$ with exactly one
source, or equivalently the number of DOAG labelled transition matrices
containing a non-zero value at coordinates~$(1,2)$.
The first line of Figure~\ref{fig:mat:decomp} illustrates the first point of the
decomposition, which yields
\begin{equation}\label{eq:Dn0}
  D_n = D_n^\star + O(v_{n-2} D_{n-1}).
\end{equation}
Note that the big-$O$ term comes from the fact that not all pairs made of a
size-$(n-2)$ variation and a size-$(n-1)$ labelled transition matrix can be
obtained this way, as discussed in the first case above.
We could actually have written~$0 \leq D_n - D_n^\star \leq v_{n-2} D_{n-1}$.

Then we decompose the matrices from~${\mathcal D}^\star$ depending of their
values on the diagonal (cases~2 to~4).
The second line of Figure~\ref{fig:mat:decomp} illustrates this decomposition.
This translates into the following identity
\begin{equation}\label{eq:Dnstar}
  D_n^\star
  = (v_{n-1}-v_{n-2}) D_{n-1}^\star
  + \frac{v_{n-1} - 2 v_{n-2} + v_{n-3}}{2} v_{n-3} D_{n-2}^\star
  + \bigO{v_{n-1} v_{n-3} v_{n-4} D_{n-3}}.
\end{equation}

Let us introduce the polynomial term~$P_n^\star$ of~$D_n^\star$ defined
by~$P_n^\star = D_n^\star / e^{n-1} \sfact{(n-1)}$.
By normalising equation~\eqref{eq:Dn0} and using equation~\eqref{eq:Pnplusk} we
have
\begin{equation*}
  P_n^\star
  = P_n + \bigO{\frac{v_{n-2}}{e (n-1)!} P_{n-1}}
  = P_n + \bigO{\frac{P_{n-1}}{n}}
  = P_n + \littleO{P_n}.
\end{equation*}
In other words, we have that~$P_n$ and~$P_n^\star$ are equivalent.
Then, by normalising equation~\eqref{eq:Dnstar} by~$e^{n-1} \sfact{(n-1)}$, we
obtain the following asymptotic expansion
\begin{align}\label{eq:Pnstar}
  P_n^\star
  = \left(1 - \frac{1}{n} + \bigO{\frac{1}{n^2}}\right) P_{n-1}^\star
  + \frac{1}{2n} \left(1 + \bigO{\frac{1}{n}}\right) P_{n-2}^\star
  + \bigO{\frac{P_{n-3}}{n^2}}\cdot
\end{align}
Since~$P_n^\star \sim P_n$ and by equation~\eqref{eq:Pnplusk}, we have
that~$\bigO{P_{n-3}n^{-2}} = \littleO{P_n^\star n^{-1}}$ and that the first term
of equation~\eqref{eq:Pnstar} dominates all the others.
As a consequence we get a refinement on our knowledge on~$P_n^\star$ (and
thus~$P_n$), that is:
\begin{equation*}
  P_n^\star \sim P_{n-1}^\star.
\end{equation*}
It is worth noting that this is the key property that makes
analysing~$P_n^\star$ possible.
From now on, we know that~$P_n^\star$ does not oscillate, and this is all
because of equation~\eqref{eq:Pnplusk}.
By re-using this new information in equation~\eqref{eq:Pnstar}, we get another
term of the expansion of~$P_n^\star$:
\begin{equation*}
  P_n^\star
  = P_{n-1}^\star \left(1 - \frac{1}{2n} + \bigO{\frac{1}{n^2}}\right).
\end{equation*}
We conclude on the asymptotic behaviour of~$P_n^\star$ using the following
classical argument.
The series of general term~$\ln\left(\!\frac{P_n^\star}{P_{n-1}^\star}\!\right)
+ \frac{1}{2n} = \bigO{n^{-2}}$ (defined for~$n \geq 2$) is convergent and,
if~$\lambda$ denotes its sum, we have that
\begin{equation*}
  \lambda - \sum_{j=2}^n \left(
    \ln\left(\!\frac{P_j^\star}{P_{j-1}^\star}\!\right)
    + \frac{1}{2j}
  \right)
  = O(n^{-1}).
\end{equation*}
Furthermore, since~$P_1^\star = 1$, we also have that
\begin{equation*}
  \sum_{j=2}^n \left(
    \ln\left(\!\frac{P_j^\star}{P_{j-1}^\star}\!\right)
    + \frac{1}{2j}
  \right)
  = \ln P_n^\star + \frac{1}{2} \left(\ln(n) + \gamma + \bigO{n^{-1}}\right)
\end{equation*}
where~$\gamma$ denotes the Euler{--}Mascheroni constant.
As a consequence, we have that
\begin{equation*}
  \ln P_n^\star + \frac{\ln(n)}{2} = \lambda + \gamma + \bigO{n^{-1}}
\end{equation*}
and thus
\begin{equation*}
  P_n^\star = \frac{e^{\lambda + \gamma}}{\sqrt{n}} \left(1 +
    \bigO{n^{-1}}\right).
\end{equation*}
By equation~\eqref{eq:Dn0}, we also get that~$P_n = P_n^\star \left(1 +
\bigO{n^{-1}}\right)$, which concludes the proof of Theorem~\ref{thm:Dnequiv},
state on page~\pageref{thm:Dnequiv} at the beginning of this section.

\subsection{Approximation of the constant}\label{sec:cstapprox}

Before concluding this section with an analysis of the behaviour of the relevant
parameters of DOAGs under the uniform model in sub-Section~\ref{sec:parameters},
we take a brief detour here to show how to estimate numerically the value of the
constant~$c$ from Theorem~\ref{thm:Dnequiv}.

Let~$D_{n,k}$ denote the number of DOAGs with~$n$ vertices (including~$k$
sources and one sink) and any number of edges.
Using the same decomposition as in Section~\ref{sec:def} and applying the same
combinatorial arguments we get
\begin{equation}
  D_{n,k}
  = \sum_{i+s \leq n-k} D_{n-1,k-1+s} \binom{s+i}{s} \binom{n-k-s}{i} i!
  = \sum_{s \geq 0} D_{n-1,k-1+s} \cdot \gamma(n-k-s,s)
\end{equation}
where
\begin{equation}
  \gamma(a,b) =\sum_{i=0}^a \binom{b+i}{b} \binom{a}{i} i!.
\end{equation}

The above sum gives an explicit way to compute~$\gamma$, but there is a
computationally more efficient way to do so using recursion and memoisation:
\begin{equation}\label{eq:gamma:rec}
  \begin{aligned}
    \gamma(a, b) &= 0 &\text{when~$a < 0$ or~$b < 0$} \\
    \gamma(0, b) &= 1 &\text{when~$b \geq 0$} \\
    \gamma(a, b) &= \gamma(a, b - 1) + a\cdot\gamma(a - 1, b) + \indicator{b = 0}
    & \text{otherwise.}
  \end{aligned}
\end{equation}
Using this recurrence formula with memoisation, the numbers~$D_{n,k}$ for
all~$n,k \leq N$ can be computed in~$O(N^3)$ arithmetic operations on big
integers, which is more efficient than using the recurrence
from~\eqref{eq:recurrence} directly.
This is expected because we eliminated the parameter~$m$.

Note that the~$D_n^\star$ sequence from Theorem~\ref{thm:Dnequiv} corresponds
to~$D_{n,1}$ and that~$D_n = \sum_{k=1}^n \hspace*{-1pt}D_{n,k}$.
Using the numbers computed by this algorithm, we plotted the first~$250$ values
of the sequences~$D_n$ and~$D_n^\star$ normalised by~$n^{-1/2} e^{n-1}
\sfact{(n-1)}$ which shows the convergence to the constant~$c$ from
Theorem~\ref{thm:Dnequiv}.
We also note that the convergence looks faster for the sequence~$D_n^\star$.
This suggests that the constant can be approximated by~$c \approx 0.4967$.
Figure~\ref{fig:plot:Dn} shows this plot as well as a zoomed-in version
near~$\frac{1}{2}$ for~$n \geq 200$.

\begin{figure}[htb]
  \centering
  \begin{tikzpicture}[xscale=0.025,yscale=4]
    \foreach \x in {0,10,...,250} \draw[black!25] (\x, 0) -- (\x, 1);
    \foreach \y in {0,0.05,...,1} \draw[black!25] (0, \y) -- (250, \y);
    \foreach \x in {0,50,...,250} {
      \draw (\x,0) node[below] {\x};
      \draw[black!50] (\x, 0) -- (\x, 1);
    }
    \foreach \y in {0,0.2,0.4,0.6,0.8,1} {
      \draw (-1,\y) node[left] {\y};
      \draw[black!50] (0,\y) -- (250,\y) {};
    }
    \draw[thick,->] (0,0) -- (255,0);
    \draw[thick,->] (0,0) -- (0, 1.05);
    \draw[Sapphire Blue,semithick] plot file {data-Dnorm.txt};
    \draw[Carmine,semithick] plot file {data-Dstarnorm.txt};
  \end{tikzpicture}
  \begin{tikzpicture}[xscale=0.125,yscale=200]
    \foreach \x in {200,202,...,250} \draw[black!25] (\x, 0.49) -- (\x, 0.51);
    \foreach \y in {0.49,0.491,...,0.51} \draw[black!25] (200, \y) -- (250, \y);
    \foreach \x in {200,210,...,250} {
      \draw (\x,0.49) node[below] {\x};
      \draw[black!50] (\x, 0.49) -- (\x, 0.51);
    }
    \foreach \y in {0.49,0.495,0.5,0.505,0.51} {
      \draw (199,\y) node[left] {\y};
      \draw[black!50] (200,\y) -- (250,\y) {};
    }
    \draw[thick,->] (200,0.49) -- (251,0.49);
    \draw[thick,->] (200,0.49) -- (200, 0.51);
    \clip (200,0.49) rectangle (250,0.51);
    \draw[Sapphire Blue,semithick] plot file {data-Dnorm.txt};
    \draw[Carmine,semithick] plot file {data-Dstarnorm.txt};
  \end{tikzpicture}
  \caption{The first values of the sequences~$\displaystyle{\frac{D_n
  \sqrt{n}}{\sfact{(n-1)} e^{n-1}}}$ in blue and~$\displaystyle{\frac{D_n^\star
  \sqrt{n}}{\sfact{(n-1)} e^{n-1}}}$ in red.}%
  \label{fig:plot:Dn}
\end{figure}

\subsection{Asymptotic behaviour of some parameters}\label{sec:parameters}

We conclude our quantitative study of DOAGs with asymptotic estimations of their
typical numbers of sources and edges under the uniform model.
The method we apply to get these results builds naturally from the methodology
developed in the rest of this section, hence illustrating the usefulness of the
matrix-based approach.

\subsubsection{Number of sources}

It follows from the previous section that the probability that a uniform DOAG of
size~$n$ has more than one source tends to zero as~$n\to\infty$.
We can refine this result and compute the probability of having a constant
number~$k$ of sources.

\begin{figure}[htb]
  \centering
  \includegraphics[scale=1]{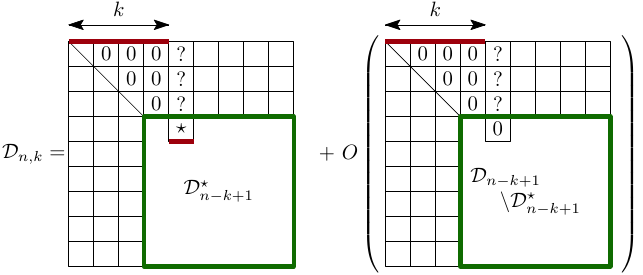}
  \caption{Decomposition of the matrices corresponding to DOAGs with~$k$
  sources}\label{fig:Dnk:decomp}
\end{figure}

We have that the number of sources of a DOAG is also the number of empty columns
in its labelled transition matrix, and that these columns are necessarily in
first positions.
Moreover, Theorem~\ref{thm:Dnequiv} gives us the intuition that most of those
transition matrices contain positive numbers near the top-left corner of the
matrix.
We thus split the set of matrices~${(a_{i,j})}_{1 \leq i, j \leq n}$ encoding
DOAGs with~$k$ sources in two categories.
\begin{description}
  \item[Case~$\boldsymbol{a_{k,k+1} > 0}$.] Intuitively, the most common
    scenario is that there is a positive entry in~$a_{k,k+1}$.
    In this case the sub-matrix~${(a_{i,j})}_{k \leq i, j \leq n}$ can be
    re-interpreted as a DOAG with only one source.
    Indeed, the condition~$a_{k,k+1}$ means that upon removing the~$(k-1)$ first
    sources of the DOAG, the decomposition algorithm does not produce any new
    source, leaving us with a single-source DOAG\@.
    We can characterise those matrices: they are made of~$(k-1)$ variations of
    size~$(n-k)$ in the first rows, and a size-$(n-k+1)$ labelled transition
    matrix corresponding to a single source DOAG below them.
    Any combination of~$(k-1)$ such variations and such a matrix can be
    obtained.
  \item[Case~$\boldsymbol{a_{k,k+1} = 0}$.] On the other hand we have the
    matrices such that~$a_{k,k+1} = 0$.
    In this case, the~$(k-1)$ first rows of the matrix are still variations of
    size~$(n - k)$.
    The lower part~${(a_{i,j})}_{k \leq i, j \leq n}$ can be seen as a DOAG with
    at least two sources because its first two columns are empty.
    Note that here, depending of the~$(k-1)$ top variations, we may have
    restriction on which DOAGs may appear in the lower part.
    For instance, if~$k=2$ and the first row is~$(0, 0, 1, 0, 0, \ldots, 0)$,
    then the DOAG of size~$(n-1)$ without any edge cannot appear in the lower
    part.
\end{description}
This dichotomy is pictured in Figure~\ref{fig:Dnk:decomp}.

From the above case analysis, we have the following bounds:
\begin{equation*}
  v_{n-k}^{k-1} D_{n-k+1}^\star
  \leq D_{n,k}
  \leq v_{n-k}^{k-1} D_{n-k+1} + \bigO{v_{n-k}^{k-1} (D_{n-k+1} -
  D_{n-k+1}^\star)}.
\end{equation*}
Thus, by virtue of Theorem~\ref{thm:Dnequiv}, we have the following estimates
when~$(n-k) \to\infty$
\begin{equation}\label{eq:dnk:bounds}
  D_{n,k} = v_{n-k}^{k-1} D_{n-k+1}^\star \left(1 + \bigO{\frac{1}{n-k}}\right),
\end{equation}
which allow us to state the following result.

\begin{theorem}[Number of sources of uniform DOAGs]
  When~$n \to\infty$ and~$(n-k)\to\infty$, we have that
  \begin{equation*}
    D_{n,k} - v_{n-k}^{k-1} D_{n-k+1} = o(D_{n,k})
  \end{equation*}
  where the little oh is uniform: it is arbitrarily smaller than~$D_{n,k}$
  when~$(n-k) \to \infty$.
  In particular for~$k$ constant, we have
  \begin{equation*}
    \frac{D_{n,k}}{D_n} \sim n^{-\binom{k}{2}}.
  \end{equation*}
\end{theorem}

\begin{proof}
  The first statement has already been established in
  equation~\eqref{eq:dnk:bounds} and the second one is straightforward to obtain
  using the equivalent~$v_n \sim e n{!}$ for the number of variations.
\end{proof}

\subsubsection{Number of edges}

Another quantity of interests of uniform DOAGs (and graphs in general) is their
number of edges.
Whereas uniform labelled DAGs have~$\frac{n^2}{4}$ edges in average, we show
here that the number of edges of uniform DOAGs is close to~$\frac{n^2}{2}$.
This has to be compared with their maximum possible number of edges which
is~$\binom{n}{2} = \frac{n(n-1)}{2}$.
This makes uniform DOAGs quite dense objects.
The intuition behind this fact is that variations have typically few zeros in
them.
Indeed, the expected number of zeros of a uniform variation is given by
\begin{equation*}
  \frac{1}{v_n} \sum_{p=0}^n p v_{n,p}
  = \frac{n!}{v_n} \sum_{p=0}^n \frac{p}{p!}
  = \frac{n!}{v_n} \sum_{p=1}^n \frac{1}{(p-1)!}
  \underset{n\to\infty}\to 1
\end{equation*}
where~$v_{n,p}$ is the number of variations of size~$n$ having exactly~$p$
zeros, which is equal to~$n!/p{!}$ by Lemma~\ref{lem:var}.
Moreover, the tail of their probability distribution is more than exponentially
small:
\begin{equation*}
  \PP_n[\text{nb zeros} \geq q]
  = \frac{n!}{v_n} \sum_{p=q}^n \frac{1}{p!}
  \underset{n,q\to\infty}=
  \frac{e^{-1}}{q!}
  \left(1 + \littleO{\frac{1}{n!}}\right)
  \left(1 + \bigO{\frac{1}{q}}\right),
\end{equation*}
where the first error term depends only on~$n$ and the second depends on~$q$ and
is uniform in~$n$.
In fact, it is straightforward to show that the distribution of the number of zeros
in a uniform variation of size~$n$ converges in distribution to a Poisson law of
parameter~$1$.

Now, recall that DOAG labelled transition matrices are a sub-class of variation
matrices, and that the number of non-zero entries in these matrices corresponds
to the number of edges of the graph.
The above discussion should make the following result intuitive.

\begin{theorem}[Number of edges of uniform DOAGs]\label{thm:edges}
  The number of edges of a uniform DOAG of size~$n$ is, in expectation,
  \begin{equation*}
    \binom{n}{2} - O(n).
  \end{equation*}
\end{theorem}

\begin{proof}
  In terms of labelled transition matrices, the theorem translates into: there
  is at most a linear number of zeros strictly above the diagonal in the matrix.
  This is what we prove here.

  For all integer~$p \geq 0$, by inclusion, we have that the number of DOAG
  labelled transition matrices with exactly~$p$ zeros strictly above the
  diagonal is upper-bounded by the number~$\VM_{n,p}$ of variation matrices with
  the same property.
  Moreover, given a vector $(p_1, p_2, \ldots, p_{n-1})$ of non-negative
  integers such that for all~$i$,~$p_i \leq i$, the number of such variation
  matrices with exactly~$p_{n-i}$ zeros in the~$i$-th line is
  \begin{equation*}
    \prod_{i=1}^{n-1} v_{i,p_i}
    = \prod_{i=1}^{n-1} \frac{i!}{p_i!}
    = \sfact{(n-1)} \prod_{i=1}^{n-1} \frac{1}{p_i!}.
  \end{equation*}
  By summation over all such vectors such that~$\sum_{i=1}^{n-1}p_i = p$, we get
  an expression for~$\VM_{n,p}$:
  \begin{equation*}
    \VM_{n,p}
    =
    \sfact{(n-1)}
    \sum_{\substack{
      p_1 + p_2 + \cdots + p_{n-1} = p \\
      \text{for all~$i$}, 0 \leq p_i \leq i
    }} \prod_{i=1}^{n-1} \frac{1}{p_i!}
    \leq
    \sfact{(n-1)}
    \sum_{\substack{
      p_1 + p_2 + \cdots + p_{n-1} = p \\
      \text{for all~$i$}, 0 \leq p_i
    }} \prod_{i=1}^{n-1} \frac{1}{p_i!}\cdot
  \end{equation*}
  In the first sum we have the constraint~$p_i \leq i$ because a variation has
  at most~$i$ zeros.
  The inequality comes from the fact that we added more terms in the sum by
  dropping this constraints.
  This allows us to interpret the sum as a Cauchy product and can express it as
  the~$p$-th coefficient of the power series~$e^x \cdot e^x \cdot \cdots \cdot
  e^x = e^{(n-1)x}$.
  It follows that
  \begin{equation*}
    \VM_{n,p} \leq \sfact{(n-1)} \frac{{(n-1)}^p}{p!}\cdot
  \end{equation*}
  As a consequence, we have the following bound for the probability that a
  uniform DOAG of size~$n$ has at most~$\binom{n}{2} - q$ zeros:
  \begin{equation}\label{eq:prob:zeros}
    \PP_n[\text{a uniform DOAG has at most~$\binom{n}{2} - q$ zeros}]
    \leq \frac{\sfact{(n-1)}}{D_n} \sum_{p \geq q} \frac{{(n-1)}^p}{p!}\cdot
  \end{equation}
  The sum in the last equation is the remainder in the Taylor expansion of
  order~$(q-1)$ of the function~$e^x$ near zero, evaluated at the point~$(n-1)$.
  By using the integral form of this remainder, we have that
  \begin{equation*}
    \sum_{p \geq q} \frac{{(n-1)}^p}{p!}
    = \int_0^{n-1} e^t \frac{{(n - 1 - t)}^{q-1}}{(q-1)!} dt
    \leq e^{n-1} \int_0^{n-1} \frac{{(n-1-t)}^{q-1}}{(q-1)!} dt
    = e^{n-1} \frac{{(n-1)}^q}{q!}\cdot
  \end{equation*}
  Furthermore, by setting~$q = \lambda (n-1)$ for some constant~$\lambda > 0$,
  and by using Stirling's formula, we get that
  \begin{equation*}
    \frac{{(n-1)}^q}{q!}
    \sim {\left(\frac{e(n-1)}{q}\right)}^q \frac{1}{\sqrt{2\pi q}}
    \sim {\left(\frac{e^\lambda}{\lambda^\lambda}\right)}^{n-1}
    \frac{1}{\sqrt{2\pi\lambda n}}\cdot
  \end{equation*}
  Finally, by using this estimate inside equation~\eqref{eq:prob:zeros}, and by
  using Theorem~\ref{thm:Dnequiv} for estimating~$D_n$, we get that there exists
  a constant~$c' > 0$ such that
  \begin{align*}
    \PP_n[\text{a uniform DOAG has at most~$\binom{n}{2} - \lambda (n-1)$ zeros}]
    &\leq \frac{\sfact{(n-1)}e^{n-1}}{D_n} \sum_{p \geq q} \frac{{(n-1)}^q}{q!} \\
    &\leq \frac{c'}{\sqrt{\lambda}}
    {\left(\frac{e^\lambda}{\lambda^\lambda}\right)}^{\mathclap{\,\,n-1}}\cdot
  \end{align*}
  The latter expression is exponentially small as soon as~$\lambda > e$ and
  dominates the tail of the probability distribution of the number of zeros
  strictly above the diagonal in DOAG labelled transition matrices, which allows
  to conclude.
\end{proof}

\section{Uniform sampling of DOAGs by vertices only}\label{sec:sampling:rej}

The knowledge from the previous section on the asymptotic number of DOAGs
with~$n$ vertices can be interpreted combinatorially to devise an efficient
uniform random sampler of DOAGs based on rejection.
Since the set of labelled transition matrices of size~$n$ is included in the set
of variation matrices of size~$n$, a possible approach to sample uniform DOAGs
is to sample uniform variation matrices until they satisfy the properties of
Theorem~\ref{thm:matrix}, and thus encode a DOAG\@.

Since the number of variation matrices is close (up to a factor of the order
of~$\sqrt{n}$) to the number of DOAGs, the probability that a uniform variation
matrix of size~$n$ corresponds to the labelled transition matrix of DOAG is of
the order of~$n^{-\frac{1}{2}}$.
As a consequence, the expected number of rejections done by the procedure
outlined above is of the order of~$\sqrt{n}$ and its overall cost is~$\sqrt{n}$
times the cost of generating one variation matrix.
Moreover, we will see that variations (and thus variation matrices) are cheap to
sample, which makes this procedure efficient.

This idea, which is a textbook application of the rejection principle, already
yields a reasonably efficient sampler of DOAGs.
In particular it is much faster than the sampler from the previous section based
on the recursive method, because it does not have to carry arithmetic operations
on big integers.
In this section we show that this idea can be pushed further using ``early
rejection''.
That is to say we check the conditions from Theorem~\ref{thm:matrix} on the fly
when generating the variation matrix, in order to be able to abort the
generation as soon as possible if the matrix is to be rejected.
We will describe how to generate as few elements of the matrix as possible to
decide whether to reject it or not, so as to mitigate the cost of these
rejections.

First, we design an asymptotically optimal uniform sampler of variations in
Section~\ref{sec:variation}, and then we show in Section~\ref{sec:fastrej}
how to leverage this into an asymptotically optimal sampler of DOAGs.

\subsection{Generating variations}\label{sec:variation}

The first key step towards generating DOAGs, is to describe an efficient uniform
random sampler of variations.
We observe that the number of zeros of a uniform variation of size~$n$ obeys a
Poisson law of parameter~$1$ conditioned to be at most~$n$.
Indeed,
\begin{equation*}
  \PP[\text{a uniform variation of size~$n$ has~$p$ zeros}]
  = \frac{v_{n,p}}{v_n} \propto \frac{\indicator{0 \leq p \leq n}}{p!}
\end{equation*}
by Lemma~\ref{lem:var}.
A possible way to generate a uniform variation is thus to draw a Poisson
variable~$p$ of parameter~$1$ conditioned to be at most~$n$, and then shuffling
a size~$p$ array of zeros concatenated with the identity permutation using the
Fisher-Yates algorithm~\cite{FY1948}.
This is described in Algorithm~\ref{algo:variation}.

\begin{algorithm}[htb]
  \caption{Uniform random sampler of variations based on the rejection
  principle.\label{algo:variation}}
  \begin{algorithmic}[1]
    \Require{An integer~$n > 0$}
    \Ensure{A uniform random variation of size~$n$}
    \Function{UnifVariation}{$n$}
      \assign{$p$}{\Call{BoundedPoisson}{$1, n$}}\label{line:callpois}
      \assign{$A$}{$[0, 0, \ldots, 0, 1, 2, \ldots, n - p]$}%
      \label{line:initarray}%
      \Comment{array of length~$n$, starting with~$p$ zeros}
      \For{$i = 0$ \textbf{to}~$n - 2$}\label{line:shuffle}
        \assign{$r$}{\Call{Unif}{$\llbracket i; n - 1 \rrbracket$}}
        \State{$A[r] \leftrightarrow A[i]$}
        \Comment{Swap entries of indices~$r$ and~$i$}
      \EndFor{}
      \return{$A$}
    \EndFunction{}
  \end{algorithmic}
\end{algorithm}

Regarding the generation of the bounded Poisson variable (performed at
line~\ref{line:callpois}), an efficient approach is to generate regular
(unbounded) Poisson variables until a value less than~$n$ is found.
Indeed, the probability~$p_n$ that a Poisson variable of parameter~$1$ is
smaller than~$n$ is
\begin{equation*}
  p_n = e^{-1} \sum_{k=0}^n \frac{1}{k!} \geq \frac{2}{e}\qquad\text{when~$n\geq1$}.
\end{equation*}
Moreover, when~$n$ is large we have~$1 - p_n \sim \tfrac 1 {(n+1)!}$.
As a consequence, the expected number of tries of a rejection procedure for
sampling conditioned $\text{Poisson}(1)$ variables is~$\frac{1}{p_n} \leq
\frac{e}{2}$.
The algorithm described by Knuth in~\cite[page~137]{Knuth1997} is suitable for
our use-case since our Poisson parameter~($1$ here) is small.
Furthermore it can be adapted to stop early when values strictly larger than~$n$
are found.
This is described in~Algorithm~\ref{algo:poisson}.

\begin{algorithm}[htb]
  \caption{Adapted Knuth's algorithm for bounded Poisson simulation}%
  \label{algo:poisson}
  \begin{algorithmic}[1]
    \Require{A Poisson parameter~$\lambda > 0$ and an integer~$n \geq 0$}
    \Ensure{A Poisson variable of parameter~$\lambda$ conditioned to be at most~$n$}
    \Function{BoundedPoisson}{$\lambda, n$}
      \Repeat{}\label{line:pois:outer}
        \assign{$k$}{$0$}
        \assign{$p$}{\Call{Unif}{$[0; 1]$}}
        \While{$(k \leq n) \wedge (p > e^{-\lambda})$}\label{line:pois1}
          \assign{$k$}{$k + 1$}
          \assign{$p$}{$p \cdot \text{\Call{Unif}{$[0; 1]$}}$}
        \EndWhile{}\label{line:pois2}
      \Until{$k \leq n$}
      \return{$k$}
    \EndFunction{}
  \end{algorithmic}
  \medskip
  \textit{NB}.\ The~$\funname{Unif}([0; 1])$ function generates a uniform real
  number in the~$[0; 1]$ interval.
\end{algorithm}

Note that this algorithm relies on real numbers arithmetic.
In practice, approximating these numbers by IEEE~754 floating points
numbers~\cite{ieee754} should introduce an acceptably small error.
Indeed, since we only compute products (no sums or subtractions), which
generally have few terms, the probability that they introduce an error should
not be too far from~$2^{-53}$ on a 64-bits architecture.
Of course this is only a heuristic argument.
A rigorous implementation must keep track of these errors.
One possible way would be to use fixed points arithmetic for storing~$p$ and to
lazily generate the base 2 expansions of the uniform variables at play until we
have enough bits to decide how~$p$ and~$e^{-\lambda}$ compare at
line~\ref{line:pois1}.
Another way would be to use Ball arithmetic~\cite{Hoeven2010, arb} and to
increase precision every time the comparison requires more bits.
The proofs of correctness and complexity below obviously assume such an
implementation.

\begin{lemma}[Correctness of Algorithm~\ref{algo:variation}]
  Given an input~$n > 0$, Algorithm~\ref{algo:variation} produces a uniform
  random variation of size~$n$.
\end{lemma}

\begin{proof}
  The correctness of Algorithm~\ref{algo:poisson} follows from the arguments
  given in~\cite[page~137]{Knuth1997}, which we do not recall here.
  Regarding Algorithm~\ref{algo:variation}, the for loop at
  line~\ref{line:shuffle} implements the Fisher-Yates~\cite{FY1948} algorithm,
  which performs a uniform permutation of the contents of the array
  \emph{independently of its contents}.
  In our use-case, this implies that:
  \begin{itemize}
    \item the number of zeros is left unchanged;
    \item given an initial array with~$p$ zeros as shown at
      line~\ref{line:initarray}, the probability to get a particular variation
      with~$p$ zeros is given by the probability that a uniform permutations
      maps its first~$p$ values to a prescribed subset of size~$p$, that
      is~$\frac{p!}{n!}$.
  \end{itemize}
  This tells us that, the probability that Algorithm~\ref{algo:variation} yields
  a particular variation with~$p$ zeros is
  \begin{equation*}
    \mathbb{P}[\funname{BoundedPoisson}(1,n) = p] \cdot \frac{p!}{n!}
    = \frac{1}{p! \sum_{k=0}^n \frac{1}{k!}} \cdot \frac{p!}{n!}
    = \frac{1}{v_n}.
    \qedhere
  \end{equation*}
\end{proof}

The minimal ``amount of randomness'' that is necessary to simulate a probability
distribution is given by its entropy.
This gives us a lower bound on the complexity (in terms of random bit
consumption) of random generation algorithms.
For uniform random generation, this takes a simple form since the entropy of a
uniform variable that can take~$M$ distinct values is~$\log_2(M)$.
This tells us that we need at least~$\log_2(v_n)$ random bits to generate a
uniform variation of size~$n$.
When~$n$ is large, we have~$\log_2(v_n) = n \log_2(n) - \frac{n}{\ln(2)} +
O(\log_2(n))$.
The uniform variation sampler we give in Algorithm~\ref{algo:variation} is
\emph{asymptotically} optimal in terms of random bit consumption: in
expectation, the number of random bits that it uses is equivalent
to~$\log_2(v_n)$.
\begin{lemma}[Complexity of Algorithm~\ref{algo:variation}]
  In expectation, Algorithm~\ref{algo:variation} performs a linear number of
  arithmetic operations and memory accesses, and consumes~$n \log_2(n) + o(n
  \log_2(n))$ random bits.
\end{lemma}

\begin{proof}
  The mean of a Poisson variable of parameter~$1$ being~$1$,
  Algorithm~\ref{algo:poisson} succeeds to find a value smaller or equal to~$n$
  in a constant number of tries in average, and each try requires a constant
  number of uniform variables in average.
  Furthermore, in order to perform the comparison~$p > e^{-1}$ at
  line~\ref{line:pois1} in the algorithm, we need to evaluate these uniform
  random variables.
  This can be done lazily, and again, it is sufficient to know a constant number
  of bits of these variables in average to decide whether~$p > e^{-1}$.

  Regarding the shuffling happening at line~\ref{line:shuffle} in
  Algorithm~\ref{algo:variation}, it needs to draw~$(n-1)$ uniform integers,
  respectively smaller or equal to~$1$,~$2$,~$3$, \ldots,~$n-1$.
  At the first order, this incurs a total cost in terms of random bits, of
  \begin{equation*}
    \sum_{k=2}^n \log_2(k) \sim n \log_2(n).
  \end{equation*}

  In total, the cost of Algorithm~\ref{algo:variation} is thus dominated by the
  shuffling, which allows to conclude on its random bits complexity.

  Regarding the number of arithmetic operations and memory accesses, generating
  Poisson variables performs in constant time using similar arguments.
  The shuffling part of the algorithm is clearly linear.
\end{proof}

Note that we count \emph{integer operations} in the above Lemma, thus
abstracting away the cost of these operations.
At the bit level an extra~$\log_2(n)$ term would appear to take into account the
size of these integers.
This type of considerations is especially important when working with big
integers as it was the case in Section~\ref{sec:def}.
However here, arithmetic operations on integers, rather than bits, seems to be
the right level of granularity as a real-life implementation is unlikely to
overflow a machine integer.

\subsection{A fast rejection procedure}\label{sec:fastrej}

Equipped with the variation sampler described above, we can now generate
variation matrices in an asymptotically optimal way, by filling them with
variations of sizes~$(n-1), (n-2), \ldots, 3, 2, 1$.
By checking afterwards whether the matrix corresponds to a valid DOAGs, and
trying again if not, we get a uniform sampler of DOAGs that is only sub-optimal
by a factor of the order of~$\sqrt{n}$.
This is presented in Algorithm~\ref{algo:rej:naive}.
This algorithm is already more efficient than a sampler based on the recursive
method, whilst naive.

\begin{algorithm}
  \caption{A simple but sub-optimal uniform random sampler of DOAGs}%
  \label{algo:rej:naive}
  \begin{algorithmic}
    \Require{An integer $n > 0$}
    \Ensure{A uniform DOAG with~$n$ vertices as its labelled transition matrix}
    \Function{UnifDOAGNaive}{$n$}
      \assign{$A = {(a_{i, j})}_{1 \leq i, j \leq n}$}{a zero-filled~$n \times
      n$ matrix}
      \Repeat{}
        \For{$i$ \textbf{from} $1$ \textbf{to} $n - 1$}
          \assign{${(a_{i, j})}_{i < j \leq n}$}{\Call{UnifVariation}{$n - i$}}
        \EndFor{}
      \Until{$A$ encodes a DOAG}\label{line:validity}
      \return{The DOAG corresponding to~$A$}
    \EndFunction{}
  \end{algorithmic}
\end{algorithm}

Checking the validity of a matrix at line~\ref{line:validity} corresponds to
checking the conditions given in Theorem~\ref{thm:matrix} at
page~\pageref{thm:matrix}.
We do not provide an algorithm for this here, as the goal of this section is to
iterate upon Algorithm~\ref{algo:rej:naive} to provide a faster algorithm and
get rid of the~$\sqrt{n}$ factor in its cost.
We will see in the following that checking these conditions can be done in
linear time.

\smallskip

\textit{Remark.}
Note that the memory footprint of a DOAG is of the order of~$n^2$ since it
typically has~$\frac{n^2}{2}$ edges as shown in the previous section, so the
number of edges might be a more natural notion of size for those objects when
talking about complexity.
If we express the complexity of Algorithm~\ref{algo:rej:naive} in terms of the number of edges~$m$
of the generating DOAG, we get that it performs~$O(m \sqrt[4]{m})$ memory
accesses and consumes~$O(m \sqrt[4]{m} \log n)$ random bits.
Under this lens, the extra~$\sqrt n$ factor incurred by the rejection is
actually only a fourth root of the more natural size parameter~$m$.

\smallskip

As we can see in Theorem~\ref{thm:matrix}, the conditions that a variation
matrix must satisfy to be a labelled transition matrix concern the shape of the
boundary between the zero-filled region between the diagonal and the first
positive values above the diagonal.
Moreover, we have seen in Theorem~\ref{thm:edges} that uniform DOAGs tend to
have close to~$\binom{n}{2}$ edges and thus only a linear number of zeros above
the diagonal of their labelled transition matrix.
We can thus expect that the area that we have to examine to have access to this
boundary should be small.
This heuristic argument, hints at a more sparing algorithm that would start by
filling the matrix near the diagonal and check its validity early, before
generating the content of the whole matrix.
This idea, of performing rejection as soon as possible in the generation
process, is usually referred to as ``anticipated rejection'' and also appears
in~\cite{DFLS2004} and~\cite{BPS1994} for instance.

To put this idea in practice, we need to implement lazy variation generation, to
be able to make progress in the generation of each line independently, and to
perform the checks of Theorem~\ref{thm:matrix} with as little information as
necessary.

\paragraph{Ingredient one: lazy variations}

\begin{figure}[htb]
  \centering
  \includegraphics[scale=1]{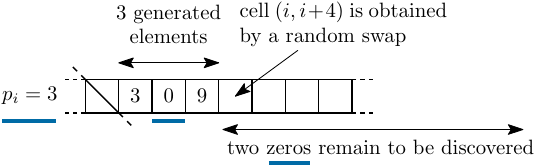}
  \caption{Ingredient one of the fast rejection-based algorithm: variations can
  be lazily generated.
  In the example, the three first elements of the variation at row~$i$ are
  known.
  When we need to generate its fourth element, we perform a swap of~$a_{i,i+4}$
  with a uniform cell of index~$j \geq i + 4$.}%
  \label{fig:ingredient:1}
\end{figure}

Fortunately, Algorithm~\ref{algo:variation} can be easily adapted for this
purpose thanks to the fact that the for loop that implements the shuffle
progresses from left to right in the array.
So a first ingredient of our optimised sampler is the following setup for lazy
generation:
\begin{itemize}
  \item for each row of the matrix (\textit{i.e.}\ each variation to be sampled), we draw
    a Poisson variable~$p_i$ of parameter~$1$ and bounded by~$(n-i)$;
  \item drawing the number at position~$(i, j)$, once we have drawn all the
    numbers of lower coordinate in the same row, can be done by selecting
    uniformly at random a cell with higher or equal coordinate on the same row
    and swapping their contents.
\end{itemize}
This is illustrated in Figure~\ref{fig:ingredient:1}.

\paragraph{Ingredient two: only one initialisation}

\begin{figure}[htb]
  \centering
  \includegraphics[scale=1]{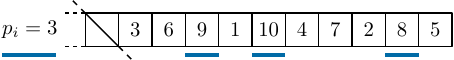}
  \caption{Ingredient two of the fast rejection-based algorithm: the zeros of
  the matrix need not be explicitly written.
  Instead, we interpret the numbers strictly larger than~$(n-i-p_i)$ as zeros.
  In this example~$(n-i) = 10$ and~$p_i = 3$ so the numbers~$8$,~$9$, and~$10$
  are seen as zeros.}%
  \label{fig:ingredient:2}
\end{figure}

A straightforward adaptation of Algorithm~\ref{algo:variation} unfortunately
requires to re-initialise the rows after having drawn the Poisson variable (see
line~\ref{line:initarray} of Algorithm~\ref{algo:variation}) at each iteration
of the rejection algorithm.
This is costly since about~$n^2 / 2$ numbers have to be reset.
It is actually possible to avoid this by initialising all the rows only once and
without any zeros.
Only at the end of the algorithm, once a full matrix have been generated, one
can re-interpret the~$p_i$ largest numbers of row~$i$, for all~$i$, to be zeros.
This is pictured in Figure~\ref{fig:ingredient:2}.

\paragraph{Ingredient three: column by column checking}

\begin{figure}[htb]
  \centering
  \includegraphics[scale=1]{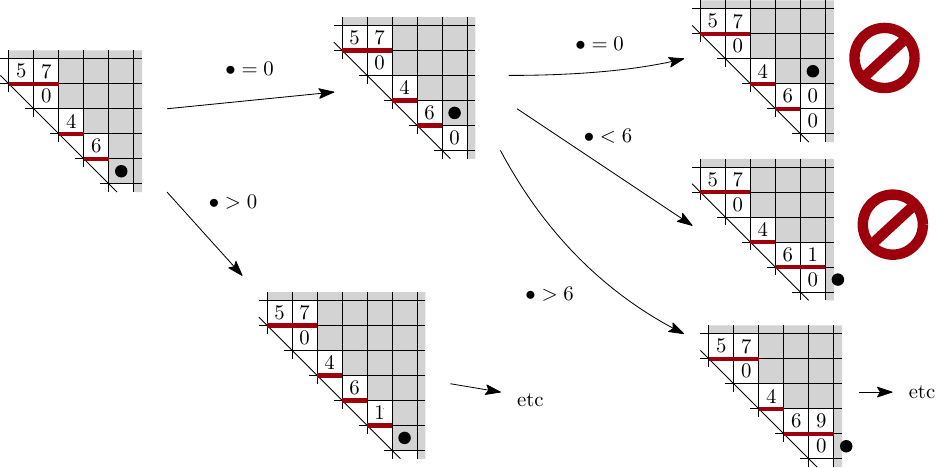}
  \caption{Ingredient three of the fast rejection-based algorithm: the
  exploration process of the cells of the matrix is dictated by the following
  algorithm.
  We proceed column by column, from bottom to top, and we advance to the next
  columns as soon as we discover a non-zero cell, or see one on our left.
  In the pictures, the bullet~$\bullet$ represent the current cell, the grey
  area represents the cells that have not yet been drawn and the thick red lines
  underline the lowest non-zero cell of each column, as before.
  Depending on the value that is drawn in the current cell, we either move up or
  to the next column.
  Whenever a non-zero cell is on our left, we can decide whether to reject or to
  keep generating.}%
  \label{fig:ingredient:3}
\end{figure}

The last detail that we need to explain is how to check the conditions of
Theorem~\ref{thm:matrix}.
As a reminder
\begin{itemize}
  \item for each~$2 \leq j \leq n$ we need to compute the number~$b_j =
    \max\,\{i \ | \ a_{i, j} > 0\}$ (or~$0$ if this set is empty);
  \item we must check whether this sequence is weakly increasing;
  \item and whenever~$b_{j+1} = b_j$, we must check that~$a_{b_j,j} <
    a_{b_j,j+1}$.
\end{itemize}
A way of implementing this is to start filling each column of the matrix from
bottom to top, starting from the column~$j=1$ and ending at column~$j=n$.
For each column, we stop as soon as either a non-zero number is found or the
constraints from Theorem~\ref{thm:matrix} are violated.
In order to check these constraints, while filling column~$j$ from
bottom~$(i=j-1)$ to top, we halt as soon as either the cell on the left of the
current cell, or the current cell is non-zero.
The case when the left cell is non-zero corresponds to when~$i=b_{j-1}$ and the
conditions of Theorem~\ref{thm:matrix} can be checked.
Recall that, per the previous point, the zero test in row~$i$ is actually~$x
\mapsto x > n - i - p_i$.
We shall prove that this process uncovers only a linear number of cells of the
matrix, thus allowing to reject invalid matrices in linear expected time.
This idea is pictured in Figure~\ref{fig:ingredient:3}.

\paragraph{The algorithm}

Putting all of this together yields Algorithm~\ref{algo:rej:opt} (on
page~\pageref{algo:rej:opt}) to generate a uniform DOAG labelled transition
matrix using anticipated rejection.
The algorithm is split into two parts.
First, the~\textbf{repeat}-\textbf{until} loop between lines~\ref{line:repeat}
and~\ref{line:until} implements the anticipated rejection phase.
At each iteration of this loop, we ``forget'' what has been done in the previous
iterations, so that~$A$ is an arbitrary matrix satisfying the following two
conditions
\begin{align}
  &i \ge j \implies a_{i,j} = 0 \\
  &\forall~1 \le i < n, \quad \{a_{i,j} \,|\, i < j \le n\} = \llbracket 1; n -
  i\rrbracket.
\end{align}
The contents of the~${(p)}_{1 \le i < n}$ vector is also forgotten and each
value is to be drawn again before any access.
The~\textbf{while} loop at line~\ref{line:while} implements the traversal of the
matrix described above: at each step, the value of the~$a_{i,j}$ is drawn and
the conditions of Theorem~\ref{thm:matrix} are checked before proceeding to the
next step.
The array~${(s_i)}_{1 \leq i \leq n}$ stores the state of each lazy variation
generator:~$s_i$ contains the value of the largest~$j$ such that~$a_{i,j}$ has
been drawn.
The second part of the algorithm, starting from line~\ref{line:for}, completes
the generation of the matrix once its near-diagonal part is known and we know no
rejection is possible any more.
This includes replacing some values of the matrix by~$0$ because of ingredient
two above.
\begin{algorithm}[htb]
  \caption{An optimised uniform random sampler of DOAGs based on anticipated
  rejection}%
  \label{algo:rej:opt}
  \begin{algorithmic}[1]
    \Require{An integer $n > 0$}
    \Ensure{A uniform DOAG with~$n$ vertices, encoded as its labelled transition
    matrix.}
    \Function{UnifDOAGFast}{$n$}
      \assign{$A = {(a_{i,j})}_{1 \leq i, j \leq n}$}%
             {the strictly upper triangular matrix~${(\indicator{j > i} \cdot (j
             - i))}_{1 \leq i, j \leq n}$}
      \assign{${(p_i)}_{1 \leq i < n}$}{uninitialised array}
      \assign{${(s_i)}_{1 \leq i < n}$}{uninitialised array}
      \Repeat{}\Comment{Anticipated rejection phase}\label{line:repeat}
        \assign{$(i, j)$}{$(1, 2)$}\Comment{position of the current cell}
        \assign{$p_1$}{\Call{BoundedPoisson}{$n-1$}}
        \While{$j \le n$}\label{line:while}
          \assign{$r$}{\Call{Unif}{$\llbracket j; n\rrbracket$}}
          \State{$a_{i,r} \leftrightarrow a_{i,j}$}
          \assign{$s_i$}{$j$}
          \If{$(a_{i,j-1} \le n - i - p_i) \wedge (a_{i,j} \not\in \llbracket
          a_{i,j-1} + 1; n - i - p_i \rrbracket)$}
            \State\textbf{break}\Comment{Rejection}
          \ElsIf{$a_{i,j} \le n - i - p_i$}
            \assign{$j$}{$j + 1$}
            \assign{$i$}{$j - 1$}
            \assign{$p_i$}{\Call{BoundedPoisson}{$1, n - i$}}
          \Else{}
            \assign{$i$}{$i - 1$}
          \EndIf{}
        \EndWhile{}
      \Until{$j > n$}\label{line:until}
      \For{$i = 1$ \textbf{to} $n-2$}\label{line:for}%
      \Comment{Completion of the matrix}
        \For{$j = i + 1$ \textbf{to} $s_i$}\label{line:for1}
          \If{$a_{i,j} > n - i - p_i$} $a_{i,j} \gets 0$\EndIf{}
        \EndFor{}
        \For{$j = s_i + 1$ \textbf{to} $n$}\label{line:for2}
          \assign{$r$}{\Call{Unif}{$\llbracket j; n\rrbracket$}}
          \State{$a_{i,r} \leftrightarrow a_{i,j}$}
          \If{$a_{i,j} > n - i - p_i$} $a_{i,j} \gets 0$\EndIf{}
        \EndFor{}
      \EndFor{}
      \return{$A$}
    \EndFunction{}
  \end{algorithmic}
\end{algorithm}

\begin{lemma}[Correction of Algorithm~\ref{algo:rej:opt}]
  Algorithm~\ref{algo:rej:opt} terminates with probability~$1$ and returns a
  uniform random DOAG labelled transition matrix.
\end{lemma}

This result is a consequence of Algorithm~\ref{algo:rej:naive} and
Algorithm~\ref{algo:rej:opt} implementing the exact same operations, only in a
different order and with an earlier rejection in the latter algorithm.
The key characteristic of this new algorithm is that is only needs to perform a
linear number of swaps in average to decide whether the reject the matrix or
not.
As a consequence it is asymptotically optimal in terms of random bits
consumption and it only performs about~$n^2 / 2$ swaps to fill the~$n \times n$
upper triangular matrix.

\begin{theorem}[Complexity of Algorithm~\ref{algo:rej:opt}]
  In average, in order to generate a uniform DOAG with~$n$ vertices,
  Algorithm~\ref{algo:rej:opt} performs~$\frac{n^2}{2} + O(n^{3/2})$ swaps in
  the matrix, and consumes~$\frac{n^2}{2} \log_2(n) + O(n^{3/2} \log_2(n))$
  random bits.
\end{theorem}

\begin{proof}
  In the rejection phase, in each column, we draw a certain number of zeros and
  at most one non-zero value before deciding whether to reject the matrix or to
  proceed to the next column.
  As a consequence, when lazily generating a variation matrix, we see at
  most~$(n-1)$ non-zero values and a certain number of zeros that we can
  trivially upper-bound by the total number of zeros (strictly above the
  diagonal) in the matrix.

  The number of variations of size~$n$ with exactly~$p$ zeros (with~$0 \le p \le
  n$) is given by~$\frac{n!}{p!}$ by Lemma~\ref{lem:var}.
  As a consequence, the expected number of zeros of a variation is given by
  \begin{equation*}
    \sum_{p=0}^n p \cdot \frac{n!}{p!} \cdot \frac{1}{v_n}
    = \left(e^{-1} + \littleO{\frac{1}{n!}}\right) \sum_{p=0}^{n-1} \frac{1}{p!}
    = 1 + \bigO{\frac{1}{n!}}.
  \end{equation*}
  It follows that the expectation of the total number of zeros of variation
  matrix of size~$n$ is~$n + O(1)$.
  This proves the key fact that, in expectation, we only discover a linear
  number of cells of the matrix in the repeat-until loop.
  Since, in expectation, we only perform~$O(\sqrt{n})$ iteration of this loop,
  it follows that we only perform~$O(n^{3 / 2})$ swaps there.
  Moreover, one swap costs~$O(\log_2(n))$ random bits, which thus accounts for a
  total of~$n^{3 / 2} \log_2(n)$ random bits in this loop.

  In order to complete the proof, it remains to show that the for loops at the
  end of Algorithm~\ref{algo:rej:opt} contribute to the leading terms of the
  estimates given in the Theorem.
  The first inner for loop at line~\ref{line:for1} replaces, among the already
  discovered values, the zeros encoded by numbers above the~$n - i - p_i$
  threshold by actual zeros.
  It is worth mentioning that this only accounts for linear number of operations
  in total, spanned over several iteration of the outer loop (at
  line~\ref{line:for}).
  The second inner for loop at line~\ref{line:for2} completes the generation of
  the matrix.
  The total number of swaps that it performs (and thus the number of uniform
  variables it draws) is~$\frac{n(n-1)}{2}$ minus the number of already
  discovered cells, that is~$n^2 / 2 + O(n)$.
  This allows to conclude the proof.
\end{proof}

By Theorem~\ref{thm:Dnequiv} on page~\pageref{thm:Dnequiv} in the previous
section, we have that~$\log_2(D_n) \sim \frac{n^2}{2} \log_2(n)$.
This shows that Algorithm~\ref{algo:rej:opt} is asymptotically optimal in terms
of random bit consumption.
Moreover, filling a~$n \times n$ matrix requires a quadratic number of memory
writes and the actual number of memory access made by our algorithm is of this
order too.

\section{Conclusion and perspectives}

In this paper, we have studied the new class of directed ordered acyclic graphs,
which are directed acyclic graphs endowed with an ordering of the out-edges of
each of their vertices.
We have provided a recursive decomposition of DOAGs that is amenable to the
effective random sampling of DOAGs with a prescribed number of vertices, edges
and source using the recursive method from Nijenhuis and Wilf.
Using a bijection with a class of integer matrices, we also have provided an
equivalent for the number of DOAGs with~$n$ vertices and designed a uniform
random sampler for DOAGs with~$n$ vertices and any number of edges.
This second sampler is asymptotically optimal, both in terms of memory accesses
and random bits consumption.

We have also showed that our approach allows to approach classical labelled DAGs
and have obtained a new recurrence formula for their enumeration.
The important particularity of this new formula is that it is amenable to
effective random sampling when the number of edges is prescribed, which was not
the case for previously known formulas.

\subsection*{Perspectives}

\paragraph{On DOAGs}

So far, we have only approached DOAGs via enumerative tools and
\textit{ad.\ hoc.}\ asymptotic techniques. A common and powerful tool in
combinatorics is the use of generating function to tackle not only asymptotic
estimation, but also the convergence in distribution of parameters, such as the
number of edges.
The book~\cite{FS2009} is a reference on this topic.
Classical approaches using ordinary, exponential, or even graphical
(\cite{PD2019}) generating functions fail in our context due to the super
factorial behaviour of the number of DOAGs.
It remains an open question whether it is possible to design a generating
function approach to our objects, which could help obtaining finer estimates,
not only over~$D_n$, but also over the low of the number of edges.

\paragraph{Multi-graph variant}

An interesting question that is left open by our work is the case of the
multi-graph variant of this model: what happens if multiple edges are allowed
between two given vertices?
This makes the analysis more challenging since there is now an infinite number
of objects with~$n$ vertices.
We thus must change our point of view and take the number~$m$ of edges into
account in addition to, or instead of, the number of vertices.
Estimating the number and behaviour of DOAGs as well as their multi-graph
counterpart, when both parameters~$n$ and~$m$ grow remains an open question and
will certainly yield very different results depending on how~$n$ and~$m$ grow in
relation to each other.
We argue that the model of multi-edge DOAGs is natural, maybe even more so than
that of DOAGs, since they encode (partially) compacted plane trees.
Quantitative aspects of tree compaction, in particular the typical compression
rate, has been studied in the past~\cite{FSS1990} in a general setting.
However, the dual point of view that consists in studying already compacted
structures directly is a more recent topic, see~\cite{EFW2021} and~\cite{GW2024}
for instance.
The class of multi-edge DOAGs generalises the classes studied in those two
papers.
Moreover, being able to sample them efficiently would give a tool to reach every
possible case (including those with double edges) when testing programs
manipulating compacted trees (such as compilers) via random generation.

Another interesting question is that of the connectivity.
We do not provide a way to count connected DOAGs directly here.
However we have already proved that, in the uniform model, they are connected
with high probability since they have only one source with high probability.
Moreover, since~$D_n$ grows extremely fast, we can also foresee that a uniform
DOAG of size~$n$ with two connected components will typically have one tiny
component of size~$1$ and a big component of size~$n-1$, and that the
asymptotic estimations of such graphs is straightforward.
This implies that sampling a uniform connected DOAG with~$n$ vertices is already
possible, and efficient, by rejection and the question of their direct
enumeration is thus mostly of mathematical interest.

\paragraph{Classical labelled DAGs}

Finally, it is also natural to wonder whether our successful approach at
efficiently sampling DOAG applies to labelled DAGs.
Of course, the asymptotics of DAGs is known~\cite{Robinson1973}.
But if a matrix encoding similar to ours is feasible, that is an encoding whose
combinatorial properties are understood well enough to avoid introduction any
bias, then it might be possible to devise an efficient, pre-computation-free,
uniform sampler for DAGs.

A starting point in this direction is the fact that uniform labelled DAGs have,
in average,~$\frac{n^2}{4}$ edges.
In terms of (upper triangular) adjacency matrices, this means that about half
the cells in the upper part of the matrix are non-zero.
This is, in a sense, much less dense than for DOAGs.
However, this is still dense enough in the sense that the DAG analogue of the
red thick path in our figures (described by the sequence~$(b_j)_{1 \leq j \leq
n}$ in Sections~\ref{sec:matrix} and~\ref{sec:sampling:rej}) can be expected to
stay close to the diagonal too.
As a consequence, the approach proposed in Section~\ref{sec:sampling:rej} for
the random generation of DOAGs is still applicable, provided we have an
efficient way to sample those paths.
Indeed, our fast-rejection procedure in Algorithm~\ref{algo:rej:opt} can be seen
as the combination of two algorithms:
\begin{enumerate}
  \item an algorithm to sample the path~$(b_j)_{1 \leq j \leq n}$ under the
    distribution induced by DOAGs;
  \item and the filling of the remaining cells of the matrix by completing the
    random variation in reach row.
\end{enumerate}
In order to design a similar approach for DAGs, we need a way to sample
the~$(b_j)_{1 \leq j \leq n}$ paths (induced by the uniform distribution on
DAGs) and the second step of the algorithm would be to fill the rest of the
matrix with Bernoulli random variables of parameter~$\frac 1 2$.
Our recent ongoing work on this topic suggests that those paths can indeed be
sampled efficiently, which will be investigated further in the near future.

Notable is that the approach presented in~\cite{KM2015} can also be seen as a
way to work with matrices while maintaining uniformity by using a combinatorial
encoding using ordered integer partitions.
A caveat however is that their approach still requires a costly pre-processing,
which we seek to avoid using a rejection-based approach.

  \subsection*{Acknowledgements}

  We would like to thank one of the anonymous referees for pointing us at
  related work that we had missed in the field of Statistics and Bayesian
  inference, and appeared to have a strong connection with our line of work.

  This research received financial support from the LIA IFUMI and the ANR PPS
  project ANR-19-CE48-0014.



\begin{thebibliography}{10}

\bibitem{BPS1994}
Elena Barcucci, Renzo Pinzani, and Renzo Sprugnoli.
\newblock The random generation of directed animals.
\newblock {\em Theoretical Computer Science}, 127(2):333--350.
\newblock \doi{10.1016/0304-3975(94)90046-9}.

\bibitem{Barnes1900}
Ernest~William Barnes.
\newblock The theory of the {{G-function}}.
\newblock {\em The quarterly journal of pure and applied mathematics},
  31:264--314.

\bibitem{BLMN2015}
Mireille Bousquet-Mélou, Markus Lohrey, Sebastian Maneth, and Eric Noeth.
\newblock {{XML}} compression via directed acyclic graphs.
\newblock {\em Theory of Computing Systems}, 57(4):1322--1371.
\newblock \doi{10.1007/s00224-014-9544-x}.

\bibitem{CEH2019}
Louis-Claude Canon, Mohamad El~Sayah, and Pierre-Cyrille Héam.
\newblock A comparison of random task graph generation methods for scheduling
  problems.
\newblock In {\em European Conference on Parallel Processing}, volume 11725 of
  {\em {{LNCS}}}, pages 61--73. Springer.
\newblock \doi{10.1007/978-3-030-29400-7\_5}.

\bibitem{CH2000}
Koen Claessen and John Hughes.
\newblock {{QuickCheck}}: A lightweight tool for random testing of {{Haskell}}
  programs.
\newblock In {\em {{ACM SIGPLAN}} International Conference on Functional
  Programming}.
\newblock \doi{10.1145/351240.351266}.

\bibitem{CMPTVW2010}
Daniel Cordeiro, Grégory Mounié, Swann Perarnau, Denis Trystram, Jean-Marc
  Vincent, and Fré­dé­ric Wagner.
\newblock Random graph generation for scheduling simulations.
\newblock In {\em 3rd International {{ICST}} Conference on Simulation Tools and
  Techniques (Simutools 2010)}, page~10. ICST.

\bibitem{CDL2024}
Julien Courtiel, Paul Dorbec, and Romain Lecoq.
\newblock Theoretical analysis of git bisect.
\newblock {\em Algorithmica}, 86(5):1365--1399, 2024.
\newblock \doi{10.1007/s00453-023-01194-0}.

\bibitem{CP2024}
Julien Courtiel and Martin Pépin.
\newblock Random generation of git graphs.
\newblock In Srečko Brlek and Luca Ferrari, editors, {\em Proceedings of the
  13th Edition of the Conference on Random Generation of Combinatorial
  Structures. {{Polyominoes}} and Tilings, Bordeaux, France, 24-28th June
  2024}, volume 403 of {\em Electronic Proceedings in Theoretical Computer
  Science}, pages 79--86. Open Publishing Association.
\newblock \doi{10.4204/EPTCS.403.18}.

\bibitem{PD2019}
Élie {d}e Panafieu and Sergey Dovgal.
\newblock Symbolic method and directed graph enumeration.
\newblock {\em Acta Mathematica Universitatis Comenianae}, 88(3):989--996.

\bibitem{DFLS2004}
Philippe Duchon, Philippe Flajolet, Guy Louchard, and Gilles Schaeffer.
\newblock Boltzmann samplers for the random generation of combinatorial
  structures.
\newblock {\em Combinatorics, Probability \& Computing}, 13(4--5):577--625.
\newblock \doi{10.1017/S0963548304006315}.

\bibitem{EFW2020}
Andrew Elvey~Price, Wenjie Fang, and Michael Wallner.
\newblock Asymptotics of minimal deterministic finite automata recognizing a
  finite binary language.
\newblock In Michael Drmota and Clemens Heuberger, editors, {\em 31st
  International Conference on Probabilistic, Combinatorial and Asymptotic
  Methods for the Analysis of Algorithms ({{AofA}} 2020)}, volume 159 of {\em
  Leibniz International Proceedings in Informatics (Lipics)}, pages
  11:1--11:13. Schloss Dagstuhl–Leibniz-Zentrum für Informatik.
\newblock \doi{10.4230/LIPIcs.AofA.2020.11}.

\bibitem{EFW2021}
Andrew Elvey~Price, Wenjie Fang, and Michael Wallner.
\newblock Compacted binary trees admit a stretched exponential.
\newblock {\em Journal of Combinatorial Theory, Series A}, 177:105306.
\newblock \doi{10.1016/j.jcta.2020.105306}.

\bibitem{Ershov1958}
Andrey~Petrovych Ershov.
\newblock On programming of arithmetic operations.
\newblock {\em Communications of the ACM}, 1(8):3--6.

\bibitem{FY1948}
Ronald~Aylmer Fisher and Frank Yates.
\newblock {\em Statistical Tables for Biological, Agricultural and Medical
  Research}.
\newblock {Oliver and Boyd}.

\bibitem{FS2009}
Philippe Flajolet and Robert Sedgewick.
\newblock {\em Analytic Combinatorics}.
\newblock Cambridge University Press.
\newblock \doi{10.1017/CBO9780511801655}.

\bibitem{FSS1990}
Philippe Flajolet, Paolo Sipala, and Jean-Marc Steyaert.
\newblock Analytic variations on the common subexpression problem.
\newblock In {\em International Colloquium on Automata, Languages, and
  Programming}, pages 220--234. Springer.

\bibitem{git2018}
Kenneth Geisshirt, Emanuele Zattin, Aske Olsson, and Rasmus Voss.
\newblock {\em Git Version Control Cookbook: {{Leverage}} Version Control to
  Transform Your Development Workflow and Boost Productivity, 2nd Edition}.
\newblock Packt Publishing.

\bibitem{GPV2021}
Antoine Genitrini, Martin Pépin, and Alfredo Viola.
\newblock Unlabelled ordered {{DAGs}} and labelled {{DAGs}}: Constructive
  enumeration and uniform random sampling.
\newblock In {\em {{XI}} Latin and American Algorithms, Graphs and Optimization
  Symposium}. Eslevier.
\newblock \doi{10.1016/j.procs.2021.11.057}.

\bibitem{Gessel1996}
Ira~Martin Gessel.
\newblock Counting acyclic digraphs by sources and sinks.
\newblock {\em Discrete Mathematics}, 160(1):253--258.
\newblock \doi{10.1016/0012-365X(95)00119-H}.

\bibitem{Gessel1995}
Ira~Martin Gessel.
\newblock Enumerative applications of a decomposition for graphs and digraphs.
\newblock {\em Discrete Mathematics}, 139(1):257--271.
\newblock \doi{10.1016/0012-365X(94)00135-6}.

\bibitem{GW2024}
Manosij Ghosh~Dastidar and Michael Wallner.
\newblock Asymptotics of relaxed k-{{Ary}} trees.
\newblock In Cécile Mailler and Sebastian Wild, editors, {\em 35th
  International Conference on Probabilistic, Combinatorial and Asymptotic
  Methods for the Analysis of Algorithms ({{AofA}} 2024)}, volume 302 of {\em
  Leibniz International Proceedings in Informatics (Lipics)}, pages
  15:1--15:13. Schloss Dagstuhl – Leibniz-Zentrum für Informatik.
\newblock \doi{10.4230/LIPIcs.AofA.2024.15}.

\bibitem{Goto1974}
Eiichi Goto.
\newblock Monocopy and associative algorithms in an extended lisp.

\bibitem{GS2005}
Radu Grose and Scott~Allen Smolka.
\newblock Monte-carlo model checking.
\newblock In {\em {{TACAS}}}, volume 3440, pages 271--286. Springer.
\newblock \doi{10.1007/978-3-540-31980-1_18}.

\bibitem{Hoeven2010}
Joris Hoeven.
\newblock Ball arithmetic.
\newblock In {\em Logical Approaches to Barriers in Computing and Complexity}.

\bibitem{Izquierdo1659}
Sebastián Izquierdo.
\newblock {\em Pharus Scientiarum}, volume~2.
\newblock sumptibus Claudii Bourgeat \& Mich. Lietard.

\bibitem{arb}
Fredrik Johansson.
\newblock Arb: Efficient arbitrary-precision midpoint-radius interval
  arithmetic.
\newblock {\em IEEE Transactions on Computers}, 66(8):1281--1292.
\newblock \doi{10.1109/TC.2017.2690633}.

\bibitem{KMSW2019}
Richard Kenyon, Jason Miller, Scott Sheffield, and David~Bruce Wilson.
\newblock Bipolar orientations on planar maps and {{SLE12}}.
\newblock {\em The Annals of Probability}, 47(3):1240--1269.
\newblock \doi{10.1214/18-AOP1282}.

\bibitem{Knuth1997}
Donald~Ervin Knuth.
\newblock {\em The Art of Computer Programming, Volume 2, Seminumerical
  Algorithms}.
\newblock Addison-Wesley Longman Publishing Co., Inc.

\bibitem{Knuth2011}
Donald~Ervin Knuth.
\newblock {\em The Art of Computer Programming, Volume 41, Combinatorial
  Algorithms, Part 1}.
\newblock Addison-Wesley Longman Publishing Co., Inc., 1 edition.

\bibitem{KM2017}
Jack Kuipers and Giusi Moffa.
\newblock Partition {{MCMC}} for inference on acyclic digraphs.
\newblock {\em Journal of the American Statistical Association},
  112(517):282--299.
\newblock \doi{10.1080/01621459.2015.1133426}.

\bibitem{KM2015}
Jack Kuipers and Giusi Moffa.
\newblock Uniform random generation of large acyclic digraphs.
\newblock {\em Statistics and Computing}, 25(2):227--242.
\newblock \doi{10.1007/s11222-013-9428-y}.

\bibitem{KSM2022}
Jack Kuipers, Polina Suter, and Giusi Moffa.
\newblock Efficient sampling and structure learning of {{Bayesian}} networks.
\newblock {\em Journal of Computational and Graphical Statistics},
  31(3):639--650.
\newblock \doi{10.1080/10618600.2021.2020127}.

\bibitem{Lecoq2024}
Romain Lecoq.
\newblock Le feu Ça brûle et l'informatique Ça bugge: Combustion et
  régression dans les graphes.

\bibitem{MDB2001}
Guy Melançon, Isabelle Dutour, and Mireille Bousquet-Mélou.
\newblock Random generation of directed acyclic graphs.
\newblock {\em Electronic Notes in Discrete Mathematics}, 10:202--207.
\newblock \doi{10.1016/S1571-0653(04)00394-4}.

\bibitem{NW1978}
Albert Nijenhuis and Herbert Wilf.
\newblock {\em Combinatorial Algorithms: {{For}} Computers and Hard
  Calculators}.
\newblock Academic Press, Inc., 2 edition.
\newblock \doi{10.1016/C2013-0-11243-3}.

\bibitem{Robinson1973}
Robert~William Robinson.
\newblock Counting labeled acyclic digraphs.
\newblock In Frank Harary, editor, {\em New Directions in the Theory of Graphs
  (Proceedings of the Third Ann Arbor Conference on Graph Theory)}, pages
  239--273. Academic Press.

\bibitem{Robinson1977}
Robert~William Robinson.
\newblock Counting unlabeled acyclic digraphs.
\newblock In {\em Combinatorial {{Mathematics V}}}, Lecture {{Notes}} in
  {{Mathematics}}, pages 28--43. Springer.

\bibitem{Robinson1970}
Robert~William Robinson.
\newblock Enumeration of acyclic digraphs.
\newblock In {\em Proceedings of the Second Chapel Hill Conference on
  Combinatorial Mathematics and Its Applications (Univ. {{North}} Carolina,
  Chapel Hill, {{NC}}, 1970), Univ. {{North}} Carolina, Chapel Hill, {{NC}}},
  pages 391--399.

\bibitem{SP1995}
Neil James~Alexander Sloane and Simon Plouffe.
\newblock {\em The Encyclopedia of Integer Sequences}.
\newblock Academic Press.

\bibitem{ieee754}
IEEE~Computer Society.
\newblock {{IEEE}} standard for floating-point arithmetic.
\newblock {\em IEEE Std 754-2008}, pages 1--70.
\newblock \doi{10.1109/IEEESTD.2008.4610935}.

\bibitem{Stanley1973}
Richard~Peter Stanley.
\newblock Acyclic orientations of graphs.
\newblock {\em Discrete Mathematics}, 5(2):171--178.

\bibitem{Stanley2011}
Richard~Peter Stanley.
\newblock Enumerative combinatorics.
\newblock {\em Cambridge studies in advanced mathematics}, 1.

\bibitem{TVK2020}
Topi Talvitie, Aleksis Vuoksenmaa, and Mikko Koivisto.
\newblock Exact {{Sampling}} of {{Directed Acyclic Graphs}} from {{Modular
  Distributions}}.
\newblock In Ryan~P. Adams and Vibhav Gogate, editors, {\em Proceedings of
  {{The}} 35th {{Uncertainty}} in {{Artificial Intelligence Conference}}},
  volume 115 of {\em Proceedings of {{Machine Learning Research}}}, pages
  965--974. PMLR.

\end{thebibliography}
\end{document}